\documentclass[12pt,preprint]{aastex}
\usepackage{apjfonts}
\usepackage{amsmath}
\usepackage{amssymb}
\usepackage{lscape}
\usepackage{subfig}
\usepackage{color}

\newcommand{\Sersic}{S\'{e}rsic\ }
\newcommand{\etal}{et~al.~}
 
\def\se#1{\S\ref{sec:#1}}
\def\Fig#1{Figure~\ref{fig:#1}} 
\def\Table#1{Table~\ref{tbl:#1}} 
\def\eg{{e.g.,~}}
\def\ie{{i.e.,~}}

\def \ion#1#2{#1{\footnotesize{#2}}\relax}
\def \hi {\ion{H}{I} }
\def \hii {\ion{H}{II}\ }
\def \magarc {mag arcsec$^{-2}$}

\def\pmb#1{\setbox0=\hbox{#1}
\kern-.025em\copy0\kern-\wd0 \kern.05em\copy0\kern-\wd0
\kern-.025em\raise.0433em\box0}
 
\slugcomment{ApJ, in press.} 

\shorttitle{Radial Migrations in Virgo Disk Galaxies}
\shortauthors{Roediger \etal 2012}

\begin{document}


\title{Stellar Populations and Radial Migrations in Virgo Disk Galaxies}

\author{Joel C. Roediger and St\'ephane Courteau}    
\affil{Department of Physics, Engineering Physics \& Astronomy, Queen's 
University, Kingston, Ontario, Canada}

\author{Patricia S\'anchez-Bl\'azquez}  
\affil{Universidad Aut\'onoma de Madrid, Deptartamento de F\'isica Te\'orica, 
E-28049 Madrid, Spain}

\and

\author{Michael McDonald}  
\affil{Kavli Institute for Astrophysics and Space Research,
Massachusetts Institute of Technology Cambridge, MA, USA}

\email{jroediger,courteau@astro.queensu.ca,p.sanchezblazquez@uam.es,\\mcdonald@space.mit.edu}


\begin{abstract}
We present new stellar age profiles, derived from well-resolved optical and 
near-infrared images of 64 Virgo cluster disk galaxies, whose analysis poses a 
challenge for current disk galaxy formation models.  Our ability to break the 
age-metallicity degeneracy and the significant size of our sample represent key 
improvements over complementary studies of field disk galaxies.  Our results 
can be summarized as follows: first, and contrary to observations of disk 
galaxies in the field, these cluster galaxies are distributed almost equally 
amongst the three main types of disk galaxy luminosity profiles (I/II/III), 
indicating that the formation and/or survival of Type II breaks is suppressed 
within the cluster environment.  Second, we find examples of statistically-significant inversions (``U-shapes'') in the age profiles of all three disk galaxy 
types, reminescent of predictions from high-resolution simulations of classically-truncated Type II disks in the field.  These features characterize the age 
profiles for only about a third ($\leq$36\%) of each disk galaxy type in our 
sample.  An even smaller fraction of cluster disks ($\sim$11\% of the total 
sample), exhibit age profiles which decrease outwards (\ie negative age 
gradients).  Instead, flat and/or positive age gradients prevail ($\geq$50\%) 
within our Type I, II and III sub-samples.  These observations thus suggest 
that while stellar migrations and inside-out growth can play a significant role 
in the evolution of \textit{all} disk galaxy types, other factors contributing 
to the evolution of galaxies can overwhelm the predicted signatures of these 
processes.  We interpret our observations through a scenario whereby Virgo 
cluster disk galaxies formed initially like their bretheren in the field but 
which, upon falling into the cluster, were transformed into their present state 
through external processes linked to the environment (\eg ram pressure 
stripping, harassment).  Current disk galaxy formation models, which have 
largely focused on \textit{field} galaxies, fail to reproduce these results, 
thus calling for adequate hydrodynamical simulations of dense galaxy 
environments if we are to understand cluster disks.  The current paper 
highlights numerous constraints for such simulations.  In the Appendix, we 
confirm the claim by \cite{Er12} that Type II breaks are absent in Virgo 
cluster S0s and discuss the detection of Type III breaks in such galaxies.
\end{abstract}



\section{INTRODUCTION}\label{sec:Intro}

The classic paradigm of disk galaxy formation posits that these systems arise 
through the dissipative collapse of rotating gas clouds \citep{FE80}.  If the 
timescales for star formation and viscosity coincide, then the resultant 
luminosity distributions within these galaxies should fall off exponentially in 
the radial direction \citep{LP87,Sl02} and stellar ages should either be 
approximately constant or decrease with galactocentric radius \citep{FC01,MD05,NO06}.  Alternatively, an exponential luminosity distribution could arise from a 
combination of supernovae feedback, variation in the angular momentum 
distribution of halo gas and inefficient star formation at low gas densities 
\citep{Du09}.  Either way, galaxies showing such surface brightness profiles 
are commonly referred to as ``Type I'' disks \citep{Fr70,BH05}.  Statistical 
studies of luminosity profiles in field galaxies (\eg \citealt{PT06}, hereafter 
PT06; \citealt{Er08}, hereafter Er08) have revealed that Type I disks in the 
field are rare (Type I fraction, $f_I \sim$ 20\%; \citealt{Gu11}, hereafter 
Gu11).  Disk galaxy luminosity profiles are instead mostly characterized by two 
exponentials (``Type II''; \citealt{vdK79}; \citealt{dG01}; \citealt{Po02}), 
the outer exponential being shallower than the inner one; the Type II fraction, 
$f_{II}$, is roughly 50\% (Gu11).  The transition between those two (nearly) 
exponential parts of a galaxy luminosity profile is referred to as the 
``break'' radius.  The opposing case, where the outer disk profile is shallower 
beyond the break is referred to as a ``Type III'' disk \citep{Er05} and occurs 
with a frequency comparable to field Type Is ($f_{III} \sim$ 30\%; Gu11).  It is 
worthwhile to note that the fractions of the three disk galaxy types within 
other environments (\ie groups, clusters) remain poorly known (but see \citealt{Ma11} and \citealt{Er12}).

The nature of Type IIIs is not yet well understood but some have speculated 
that their breaks represent either the transition from the disk to an outer 
spheroid \citep{Er05} or the aftermath of minor mergers onto gas-rich disks 
which promoted stars from the progenitors to the remnants' outskirts 
\citep{Yo07}.  Conversely, the range of morphology in spiral disks with Type II 
breaks in PT06 suggests that a variety of mechanisms can generate such breaks 
in disk galaxies, yet only that pertaining to truncated star-forming disks, for 
so-called classical truncations (hereafter ``Type II-CT'' disks), has been 
studied in earnest thus far.  The ensuing discussion provides a broad overview 
of our current knowledge about the formation of such galaxies.

The luminosity profile breaks in Type II-CTs have historically been attributed 
to either angular momentum thresholds during the formation of disk galaxies 
\citep{vdK87,vdB01}, angular momentum redistribution via subsequent secular 
evolution of assembled disks \citep{De06,Bo07,Fo08}, or a density-dependent 
star formation law \citep{Ke89,EP94,Sc04}.  Although the redistribution 
scenario can account for the large scatter in the age-metallicity plane for 
stars within the solar neighbourhood \citep{Ed93,Ha08} (\ie chemical evolution 
models without redistribution fail to reproduce the observed scatter; \citealt{Wi96}; \citealt{SB02}), the fact that UV emission is frequently detected beyond 
the optical extent of many disk galaxies (including Type IIs; \citealt{GdP05}; 
\citealt{Th07}) strongly suggests that in-situ star formation contributes on 
some level to building the stellar mass in their outer regions.  Thus it is 
quite likely that some hybrid of the above scenarios (\eg angular momentum 
redistribution plus density-dependent star formation) is likely responsible for 
the existence of Type II-CT disks.

Various models have been proposed to explain the redistribution of angular 
momentum that may assist in the formation of Type II-CT disks.  One such model 
attributes this redistribution to an increase in the random velocities of disk 
stars caused by gravitational interactions with molecular clouds \citep{SS53,Wi77,Wi96}, but it fails to reproduce the aforementioned scatter in the Galactic 
age-metallicity relation \citep{No04}.  Alternatively, individual disk stars 
can have their angular momentum redistributed by changing the guiding centers 
of their orbits.  So-called ``radial migrations'' of stars can be produced by a 
number of mechanisms, the first being postulated by \cite{SB02} in the context 
of resonant scatterings of disk stars off of transient spiral density waves.  
Given a superposition of several spiral patterns, Sellwood \& Binney showed 
that significant displacements ($\lesssim$ 4 kpc) of guiding centers could be 
achieved.  \cite{MF10} further argued that an overlap of bar and spiral arm 
resonances could drastically enhance the efficiency of migration within a Milky 
Way-like disk, mixing it within $\sim$3 Gyr \citep[see also][]{De06}.  Lastly, 
\cite{Qu09} showed that radial migrations of stars to the outskirts of galaxy 
disks could arise via tidal perturbations during the pericenter passages of 
dwarf satellites (\eg Sgr dSph; see also \citealt{Bi12}).  Although its 
efficiency remains to be quantified, this mechanism fits quite naturally within 
the current cosmological paradigm.  Overall, and given the many mechanisms 
through which radial migrations can be incited, can observations of external 
galaxies distinguish between them?

Sellwood \& Binney's resonant scattering mechanism was revisited by \cite{Ro08} 
(hereafter R08) through their high-resolution simulation of an idealized galaxy 
disk which included both gas and star formation.  Their disk rapidly developed 
a Type II-CT break that persisted over several dynamical times, the origin of 
which was linked to the combination of angular momentum conservation within the 
collapsing gas cloud, a surface density threshold in their star formation 
recipe and radial migrations populating the region beyond the break with inner 
disk stars.  Given the inside-out growth of the inner disk and that the 
diffusion length for stars scales with the square-root of their age, R08 found 
that the stellar age profile of their simulated disk was described by both a 
negative and positive gradient interior and exterior to its break, 
respectively.  In the wake of that study, \cite{MS09} (hereafter MS09) and 
\cite{SB09} (hereafter SB09) investigated the formation of Type II-CT breaks in 
the context of \textit{cosmological} simulations.  While these authors also 
found ``U-shaped'' age profiles for their model disks associated with rapid 
decreases in the gas densities beyond their break radii, SB09 showed that the 
decrease in their simulation arose from the onset of a warp in the gas disk at 
the break rather than an angular momentum threshold (\ie it is physically well-motivated).  It is also worth stressing that the above simulations focussed on 
relatively isolated systems; comparable simulations for cluster galaxies are 
currently lacking.

While the above simulations of Type II-CT disk formation uniformly suggest that 
these galaxies are described by U-shaped stellar age profiles, such radial age 
inversions have only been explicitly detected in three Type II galaxies so far 
(NGC 2684, NGC 6155 and NGC 7437) by \cite{Yo10,Yo12}, while the age profiles 
for the remaining Type IIs in their sample which could be measured beyond their 
respective breaks (IC 1132, NGC 4904, NGC 6691) do not exhibit a significant 
increase with radius in those regions.  Furthermore, amongst their positive 
detections, these authors found that the inversion occurs well within the break 
in all cases.  Other detections of inverted age profiles have also been 
achieved through studies of the resolved stellar populations of three nearby 
Type II galaxies (M33, NGC 4244 and NGC 7793; \citealt{dJ07}; \citealt{Wi09}; 
\citealt{Ba11}; \citealt{RS12}), albeit by somewhat indirect means, such as 
surface brightness profiles for different age groups of stars or distributions 
of stellar age/star formation rate for several HST fields.  \cite{Ba08} and 
\cite{Az08} have both claimed age inversion detections in larger samples of 
nearby and distant Type II galaxies, respectively, given that their colour 
profiles invert at their break radii.  These findings, however, were 
benchmarked on optical colours, the translation of which into ages is frought 
with uncertainty due to the well-known age-metallicity degeneracy.  Signal-to-noise constraints also required stacked colour profiles rather than those of 
individual galaxies (as we do here).  Lastly, in their stellar population 
analysis of S0 galaxies, \cite{PC11} found that nearly a quarter of their 
sample exhibited very old ages in their outskirts, consistent with the above 
predictions and detections of U-shaped age profiles.

Despite their intrinsic uncertainties, stellar population data are in principle 
ideal to test the predictions of existing disk galaxy formation models.  Such a 
test represents a key step towards a definitive picture of many observable 
properties of disk galaxies, including Type II-CT breaks and whether they arise 
from a density-dependent star formation law, angular momentum thresholds or 
redistribution, or some combination thereof.  The intent of this work is to 
present such a test in the context of all three types of disk galaxies by 
taking advantage of our recent analysis of both the luminosity and stellar 
population profiles of Virgo cluster galaxies \citep{MD11,Ro11b}.  Although 
hydrodynamic models of disk galaxies which include radially-resolved stellar 
age information only exist for field environments\footnote{Hydrodynamic models 
of cluster galaxies have thus far mostly addressed the study of ram-pressure 
stripping; \citealt{Ab99}; \citealt{Vo01}; \citealt{RH05}; \citealt{MC08}.}, 
comparing their predictions against our data can still be insightful towards 
fashioning an understanding of the role of environment on the evolution of 
these galaxies (\ie nature versus nurture) and constraining future models of 
cluster evolution.  Specifically, we aim to increase the number of explicit 
detections of U-shaped age profiles in all disk galaxy types, thereby 
bolstering the empirical evidence that radial migrations are a generic aspect 
of galaxy evolution, whilst simultaneously drawing attention to cases of 
non-detections and the very plausible notion that other evolutionary processes 
(perhaps tied to environmental effects, such as ram pressure stripping) may 
either prevent or obliterate age inversions in galaxies.  The latter goal is 
made all the more imperative given the intriguing results achieved by 
\cite{Yo10,Yo12} described above.  To our knowledge, the present work is a 
first attempt towards explicitly studying stellar populations as a function of 
disk galaxy type in the context of current generation galaxy formation models 
and within the cluster environment.

We highlight in \se{Data} various aspects of our database of Virgo galaxies' 
luminosity and stellar population profiles which pertain to this study and 
present, in \se{R&D}, our data analysis in the context of Type I, II and III 
disk galaxy formation models.  Conclusions are presented in \se{Conc}.  We 
examine in the Appendix the recent claim by Erwin \etal (2012; hereafter Er12) 
that cluster S0 galaxies do not exhibit Type II breaks.


\section{DATA}\label{sec:Data}

Our investigation into the origins of the three principal disk galaxy types in 
dense environments stems from our recent analysis of the radially-resolved 
colours and stellar populations of a complete, magnitude-limited sample of 283 
Virgo cluster galaxies spanning all manner of morphological types and surface 
densities \citep{Ro11a,Ro11b}.  This analysis was facilitated by the 
availability of optical ($gri$) and near-infrared ($H$) luminosity profiles for 
the aforementioned sample from \cite{MD11}.  In \cite{Ro11b}, we appealed to 
stellar population models corresponding to an exponential star formation 
history of variable timescale and a broad range of metallicities to 
homogeneously convert those optical-infrared luminosity profiles into estimates 
of the underlying mean stellar age and metallicity profiles for all 283 
galaxies.  These luminosity, age and metallicity profiles thence enabled the 
present work.  The reader may consult \cite{MD11}\footnote{\tt http://www.astro.queensu.ca/$\sim$virgo} for the presentation of our luminosity profiles and 
\cite{Ro11b} for further details about our adopted stellar population models 
and the derivation of our age and metallicity profiles.

Our Type I, II and III sub-samples are drawn from the 85 disk galaxies in the 
Virgo cluster survey of \cite{MD11}.  Their morphological types were taken from 
the NASA Extragalactic Database\footnote{\tt http://ned.ipac.caltech.edu} (NED; 
\ie Sa$-$Sm, Im morphologies).  For reasons described in the Appendix, the 55 
S0 galaxies in the McDonald \etal sample were left out of this analysis.  Of 
the initial sample of 85 disk galaxies, we discarded fourteen on account of 
severe dust extinction (which would skew their stellar population diagnostics), 
another four due to clear signs of strong overlap/interaction with neighbouring 
systems or contamination by a bright foreground star/\hii\ region (yielding 
artificial Type III luminosity profiles), and three others which were too faint 
for their stellar populations to be reliably modelled.  These various cuts 
leave us with a sample of 64 bona fide, largely non-interacting Virgo cluster 
disk galaxies with low to moderate attenuation and high-quality 
optical-infrared luminosity profiles, in order to analyse both the broad 
structures and stellar populations of their disks.

The $griH$ luminosity profiles for each of our disk galaxies were inspected 
visually by the authors and classified according to their Type I/II/III 
features and following the detailed schemes of PT06 and Er08 (but see footnote 
7 in \se{R&D-TypeIIs}.).  Specifically, a galaxy was assigned a Type I, II or 
III class if its luminosity profile could be adequately described by a single 
exponential drop-off from the intermediate to outer regions of the galaxy (Type 
I), or exhibits a sharp break from an inner exponential section to either a 
steeper (Type II), or yet a shallower (Type III) one at some common radius 
amongst all bands.  If the aforementioned break is not seen at all bands, 
\textit{including} the near-infrared, or if it occurs in a section of the 
luminosity profile that is dominated by sky errors (where the profile point-to-point fluctuations are comparable to sky flux errors), then no physical or 
statistical basis exists for claiming that the associated galaxy be classified 
as anything other than a Type I.  On that basis, many of our galaxies are 
better classified as Type Is rather than IIIs since profile breaks in the 
latter typically occur at low surface brightnesses (27-28~\magarc at $i$-band) 
where sky errors are dominant.  Fortunately, these issues do not affect our 
reported Type II disks. 

Altogether, we find that our sample of Virgo cluster disk galaxies consists of 
$\sim$20 Type Is, 22 Type IIs and $\sim$22 Type IIIs\footnote{The Type I and 
III numbers are somewhat uncertain since deeper imaging for these galaxies 
\citep{Fe12} could call for a type revision.}.  \Fig{Images-TypeII} shows the 
$i$-band images for our Type II sub-sample, where the ellipses denote the 
locations of the breaks (red) and probable extents of the bars (green; if 
present, as gauged by eye) in these systems, while in \Table{TypeFrctns} we 
present the fractions of our Type I, II and III sub-samples contributed by each 
of the principal disk morphological types (Sa$-$Sb, Sbc$-$Sm and Im), 
normalized to our total disk sample (\ie the sum of all of the fractions is 
unity).  The morphological breakdown in \Table{TypeFrctns} shows that our Type 
I and III sub-samples are largely comprised of late-type disks and irregulars, 
and early- and late-type disks, respectively, while our Type II sub-sample 
contains a range of morphologies.  Conversely, \Table{TypeFrctns} also shows 
that our early-type disks are mostly described by Type III profiles, while our 
late-type disks and irregulars are comprised of nearly-equal fractions of Type 
I$-$III and Type I$-$II profiles, respectively.  These results conflict with 
Gu11, Er08 and PT06 who found that 28\%/27\%/10\%, 21\%/48\%/60\% and 
51\%/24\%/30\% of S0$-$Sb/SB0$-$SBb/Sb$-$Sdm galaxy types possess Type I, II 
and III profiles, respectively, and that the fraction of Type Is and IIs per 
morphological bin decreases and increases from early to late galaxies, 
respectively, while that for Type IIIs varies between 20\% (Sa, Sd, Sm) and 
50\% (S0, Sb).  \textit{The resolution of this conflict most likely lies in the 
fact that PT06, Er08 and Guti{\'e}rrez \etal targeted field galaxies, whereas 
our sample is drawn from a cluster environment}.  Indeed, Er12 recently 
reported a complete dearth of Type II profiles amongst Virgo S0s, while \cite{Mo99} showed that multiple fly-bys between cluster galaxies (so-called 
``harassment'') can transform Type I disks into Type IIIs.  These evidences, 
paired with our Type I/II/III statistics, provide initial support for the main 
theme of the present paper: that the evolution of disk galaxies is 
substantially (and verifiably) different in field versus cluster environments 
(see \S~5 in Er12).  This conclusion should stand even if deeper optical 
imaging for larger samples of Virgo disk galaxies \citep{Fe12} should reveal an 
increase in the number of Type III systems within this cluster (at the expense 
of the Type I fraction), as it is truly the relative absence of Type II disks 
in Virgo that stands out.

We present in Figures~\ref{fig:SBPrfs-TypeI}-\ref{fig:SBPrfs-TypeIII} the 
$griH$ surface brightness profiles of our Type I, II and III systems ($g$ = 
blue, $r$ = green, $i$ = orange, $H$ = black).  The Virgo Cluster Catalog (VCC) 
galaxy numbers from \cite{Bi85}, detailed morphological classifications from 
NED, and, for our Type IIs alone, specific Type class (\eg II-o.CT) have been 
included in all panels, as well as the break locations for our Type II and III 
sub-samples (red vertical lines).  We also show in each panel the 1$\sigma$ sky 
error envelope of the $i$-band dataset (black continuous line) to illustrate 
the importance of such envelopes in evaluating the behaviour of disk galaxies' 
luminosity profiles in their outskirts: again, if the extrapolation of the 
inner (mostly exponential) profile to the galaxy's outskirts falls within its 
sky error envelope, then no statistical basis exists for claiming that that 
disk is anything but a Type I.  Other cautionary cases also exist.  For 
instance, the apparent Type II breaks found at 30\arcsec in the profiles for 
VCC 1326 and 1811 (both Type Is) are due to the isophotal contours intercepting 
the edge of a bar and spiral arm at these locations, respectively, while the 
apparent Type III break at 25\arcsec in the profiles for VCC 1508 (a Type II) 
most likely marks the transition from its bulge to its disk.  This and many 
other galaxies in our Type II sub-sample do appear to host Type III breaks 
further out in their disks however, but we nevertheless retain them in that sub-sample (and simply identify them as Type II+IIIs) as we suspect that they fell 
into the cluster as Type IIs but developed their Type III breaks via harassment 
\citep{Mo99}.  The existence of these exceptional cases reinforces the notion 
that classification of disk galaxy luminosity profiles remains a somewhat 
subjective process, and as such, must be performed and interpreted with care.

As to the role that sky error envelopes play in determining a disk galaxy's 
type, we examine in our Appendix the recent claim by Er12 that roughly half of 
all bright Virgo S0s possess Type III disks.  It will be found that, at least 
for those galaxies where our respective samples overlap, many of Er12's Type 
III detections depend critically on the assumed sky level and could be labeled 
as pure \Sersic systems with $n>1$.  Still, their reported lack of Type II 
systems and existence of at least some disk upturns (Type IIIs) in cluster S0 
galaxies appears to be genuine.


\section{RESULTS \& DISCUSSION}\label{sec:R&D}
\subsection{Type Is}\label{sec:R&D-TypeIs}
The mean stellar age profiles of our 20 Type I galaxies are shown in \Fig{APrfs-TypeI}, where galactocentric radii have been expressed in terms of both 
physical (kpc; bottom axis) and effective (r$_e$; top axis) units, assuming a 
distance of 17 Mpc to the Virgo cluster \citep{Me07}.  We find that the 
majority of these galaxies are characterized by age gradients (change in age 
per unit radius) which are either positive or flat (40\% and 35\% of our Type I 
sub-sample, respectively).  This contrasts with most Type I formation models 
which predict that these galaxies form from inside-out \citep{FC01,MD05,NO06,Du09}, implying that they should harbour \textit{negative} age gradients, whereas 
such gradients are only found in two of our Type Is (10\%).  Some nuances in 
these models actually allow for flat \citep{FC01} or positive \citep{Ro04} age 
gradients within purely exponential disks depending on whether star formation 
begins during or after the assembly of the gas disk, or if gas-rich minor 
mergers/accretion events occur at late epochs, respectively.  Unfortunately, 
high-resolution hydrodynamic models to tackle the case of disk galaxy formation 
and/or evolution within the cluster environment are still lacking.  Based on 
the above observations, one then wonders if existing models (largely geared 
towards isolated galaxies and most often predicting negative age gradients) can 
reproduce the high frequency of positive and flat age gradients when cast in 
the context of denser galaxy environments.

Short of appropriate models to compare our observations, we may speculate that 
the preponderance of flat and positive age gradients in Virgo Type Is is an 
artifact of the galaxy environment; cluster galaxies experience a variety of 
physical processes that field galaxies do not.  For instance, by removing the 
gas from and thereby quenching star formation in the outskirts of a galaxy 
disk, it is conceivable that ram pressure stripping via the inter-cluster 
medium could evolve a negative age gradient into either a flat or positive 
one.  The two possible outcomes would depend upon the galaxy's ability to 
retain a central gas reservoir thus maintaining star formation there.  Our 
knowledge of these galaxies' neutral gas deficiencies, Def$_{\hi}$\footnote{Def$_{\hi}$ is a popular metric for measuring the degree of ram pressure stripping 
suffered by cluster galaxies; \citealt{HG84}} remains incomplete however, while 
the Type Is with flat or positive age gradients whose Def$_{\hi}$ values are 
known \citep{Ga05} a range of neutral gas deficiencies $>$ 1 dex.  This seems 
inconsistent with our above hypothesis.  Since Def$_{\hi}$ measures the atomic 
gas content of cluster disks relative to their field (isolated) counterparts, 
galaxies having more positive Def$_{\hi}$ values would have presumably lost 
proportionally more of their cold gas to ram pressure than systems with more 
negative Def$_{\hi}$ values and, therefore, should host more evolved stellar 
populations throughout their volumes.  Galaxy-galaxy or galaxy-cluster 
harassment \citep{Mo96,Mo99}, on the other hand, may offer an attractive 
alternative, at least for flat age gradients, if the tidal forces can promote 
effective mixing of stellar disks.  We revisit this issue in \se{R&D-Enviro}.

Even more tantalizing than the (unexpected) high incidence of flat and positive 
age gradients in Virgo Type Is is the existence of U-shaped age profiles in the 
remaining three (15\%) galaxies from this sub-sample (VCC 0583, 1410 and 1566 
in \Fig{APrfs-TypeI}), akin to those found in simulations of field Type II-CT 
disks (R08; MS09; SB09).  The hallmarks of such age profiles -- their minima -- 
are identified (where they occur) in \Fig{APrfs-TypeI} by the blue vertical 
lines.  Although the location of each age minimum was determined by eye, its 
manifestation was only deemed legitimate if the age gradient interior and 
exterior to it was computed to be negative and positive, respectively.  That 
is, an age profile minimum is deemed meaningful only if a statistically genuine 
U-shape can be defined for the profile.  With this in mind, we measured the age 
gradients interior and exterior to each of the identified minima in \Fig{APrfs-TypeI} (via bootstrap linear least-squares fitting) and found that they all 
satisfy the above stipulations and, most importantly, they are statistically 
different from zero.  To avoid the effects of atmospheric blurring in our 
interior age gradient measurements, we neglected the centralmost data point 
from each of the galaxy age profiles; in \cite{Ro11a} we showed that, for each 
galaxy in our sample, this data point encloses the maximum seeing disk.  The 
age gradients and locations of age profile minima (r$_{\text{Min}}$, where 
applicable) for our Type I sub-sample are presented in Tables \ref{tbl:AgeGrads}-\ref{tbl:TypeIDat}, respectively, while the fractions of these galaxies which 
exhibit U-shaped age profiles, or not, are shown in \Table{AGrdFrctns}, for the 
sake of completeness.  No error on the inner age gradient for VCC 0583 is 
quoted because the number of data points interior to its age minimum was too 
low to bootstrap the profile fit in that region; the value of that gradient for 
VCC 0583 in \Table{AgeGrads} was instead obtained via a simple linear least-squares fit.

Our discovery of statistically-significant inversions in the age profiles in 
three members of our Type I sub-sample challenges existing formation models for 
these galaxies; the latter models have never predicted such stellar population 
trends.  Although sites of recent star formation in disk galaxies (\eg spiral 
arms) could presumably depress the mean stellar ages in those regions, such 
effects, on their own, should not create the kind of \textit{global} minima 
that we have found in these galaxies' age profiles.  Moreover, ours is not the 
first detection of such peculiar behaviour in the stellar populations of Type I 
disks.  For instance, \cite{Vl09} found that the stellar metallicity profile 
for the canonical (field) Type I disk galaxy, NGC 300, inverts from a negative 
gradient to a positive one, analogous to our findings for our three peculiar 
Type Is, albeit at a much larger galacto-centric radius (10 kpc) and in terms 
of metallicity rather than age.  Guided by simulations which suggest that U-shaped radial distributions of stellar population properties in disk galaxies are a 
telltale sign of stellar migrations, we might speculate that these Virgo Type 
Is represent a sub-population of Type IIs whose outer disks were fully 
populated by a very efficient migration episode.  Clearly though, the existence 
of this peculiar galaxy population ought to motivate further exploration of 
Type I galaxy formation models.

\subsection{Type IIs}\label{sec:R&D-TypeIIs}
The mean stellar age profiles of our 22 Type II galaxies are shown in \Fig{APrfs-TypeII}.  We find that the behaviours of these galaxies' profiles are most 
often described by either positive age gradients or statistically-significant U-shapes (each amounting to 36\% of the sub-sample\footnote{Unfortunately, the 
luminosity profiles for VCC 664, 1555 and 1952 lacked sufficient flux to enable 
a measurement of their ages either up to (664, 1952) or much beyond (1555) 
their breaks, so that the fraction of U-shaped age profiles in our Type II sub-sample may be as high as 45\%.}; \Table{AGrdFrctns}), where the latter are 
reminiscent of predictions from current models of Type II-CT disk formation.  
The locations of the breaks and age minima in our Type II systems are shown 
(where applicable) by the red and blue ellipses in \Fig{Images-TypeII} and the 
orange and blue vertical lines in \Fig{APrfs-TypeII}, respectively.  As for our 
Type I sub-sample, detections of U-shaped age profiles were only deemed 
legitimate if the age gradients interior and exterior to each of the minima 
identified in \Fig{APrfs-TypeII} were measured via bootstrap linear fitting to 
be negative and positive, respectively, and statistically different from zero.  
In all identified cases, the fitted gradients agree with these criteria, as 
seen from \Table{AgeGrads}.  Although the central data points from these 
galaxies' inner age profiles were typically neglected during this procedure, we 
preserved them for VCC 1021 and 1654 in order to measure their inner age 
gradients; as before though, no bootstrap errors are reported for them in 
\Table{AgeGrads}.  While these two galaxies' inner gradients should be treated 
as upper limits, their specific values do not affect our conclusions.

We summarize in \Table{TypeIIDat} pertinent information for our Type II sub-sample, including the locations of their breaks (r$_{\text{Brk}}$) and age profile 
minima (r$_{\text{Min}}$; where applicable), both of which are listed in terms of 
physical units (kpc) and relative to these galaxies' $H$-band effective radii, 
r$_{\text{e}}$, from \cite{MD09}.  The profile breaks are located within the 
effective radii for half of this sub-sample, while the remainder of our Type 
IIs exhibit their breaks mostly within the 1-2 r$_{\text{e}}$ annulus.  In terms 
of physical units, however, the distribution of break radii seems bimodal, with 
these galaxies clustering about r$_{\text{Brk}}$ < 2 kpc or r$_{\text{Brk}}$ = 5-6 
kpc.  For those Type IIs with U-shaped age profiles, we find that they do not 
distinguish themselves from the whole sub-sample in terms of either 
r$_{\text{Brk}}$ or any of the non-parametric quantities measured by \cite{MD09} 
(\ie concentration, luminosity, surface brightness, size).  Moreover, while 
their distribution of r$_{\text{Min}}$ is concentrated within 1.5 r$_{\text{e}}$, it 
is spread over a much wider range in terms of physical units ($\sim$0.5-8.5 
kpc).  Thus this special sub-population of Virgo Type IIs cannot be linked to 
any specific disk galaxy structural properties.

\Table{TypeIIDat} also provides the morphologies and detailed luminosity 
profile classifications of our Type II galaxies, from which we infer that the 
Type II phenomenon amongst cluster galaxies is rarely associated with a bar.  
This should not necessarily be interpreted as suggesting that a bar has little 
effect on these galaxies' outskirts though, as a significant fraction of our 
Type II sub-sample bears morphological features which are consistent with 
previous bar activity (\ie rings, lenses).  Moreover, many of the latter 
systems exhibit U-shaped age profiles.  The detailed profile classifications in 
\Table{TypeIIDat} were determined following PT06's and Er08's scheme, whereby 
the designations ``i'' and ``o'' refer to the break occurring interior or 
exterior to the bar (if present), while ``AB'' and ``OLR'' indicate whether the 
break is related to either strong $m=1$ outer-disk asymmetries or an outer 
Lindblad resonance (\ie ring), respectively.  If both the ``AB'' and ``OLR'' 
designations may be ruled out for any given galaxy, it is deemed to possess a 
classical truncation (``CT'')\footnote{It should be noted that the Type II 
classification scheme created by PT06 and Er08 was intended to provide a 
framework under which the origin(s) of such galaxies could be understood; that 
is, it is largely hypothetical.  For instance, whether the breaks in Type 
II-o.OLR disks actually occur at their outer Lindblad resonances has yet to be 
explictly demonstrated vis-{\' a}-vis measurements of bar pattern speeds.  Thus 
our detailed classifications of the breaks in Virgo Type IIs should be 
interpreted with caution.}.  The detailed classifications for our Type IIs show 
that this phenomenon within the cluster environment is most often attributed to 
an asymmetry in the disk ($f_{AB}$ = 0.5), while classical truncations account 
for just under a third ($f_{CT}$ = 0.33) of the Type II breaks we detect.  The 
distribution of detailed Type II classes amongst Virgo disks also appears to be 
at odds with observations for field disks (albeit for late-type morphologies 
alone), for which most Type II breaks represent classical truncations ($f_{CT}$ 
= 0.5), followed by disk asymmetries and resonances in roughly equal 
proportions ($f_{AB/OLR}$ = 0.2; PT06).  This discrepancy, however, may also be 
accounted for by the nature of our sample since cluster galaxies have likely 
suffered tidal interactions more often throughout their evolution than field 
galaxies (and would therefore host asymmetric disks), which is to suggest that 
dense galaxy environments preferentially favour the development of different 
Type II classes (unlike findings for field spirals).

For those Type II galaxies in our sub-sample which exhibit U-shaped age 
profiles, we wish to compare the locations of their age minima against those of 
their breaks, given that simulations of Type II-CT disks in the field suggest 
that these locations should coincide (\ie the luminosity profile break and age 
profile inversion phenomena are causally-connected).  We plot in 
\Fig{rMinvsrBrk} the bivariate distribution of r$_{\text{Min}}$ against 
r$_{\text{Brk}}$ for the eight Virgo Type IIs which exhibit statistically-significant age minima, where the dashed line is a one-to-one relation.  For simplicity, 
we have plotted the two sets of locations in terms of these galaxies' $H$-band 
effective radii.  The points are coloured according to galaxy morphologies 
(orange = Sa$-$Sbc, green = Sc$-$Sm, blue = Im), while the point types 
correspond to detailed Type II classifications (triangle = i, circle = CT, 
square = AB, cross = OLR), with open and solid points denoting barred and 
unbarred systems, respectively.

\Fig{rMinvsrBrk} shows a clear correlation between the breaks and age minima 
locations for Type II galaxies in the Virgo cluster.  A bootstrap linear fit to 
the data yields a Pearson correlation coefficient of $r$ = +0.92 and a slope of 
+1.59 $\pm$ 0.30.  The existence of such a correlation, never reported before, 
expands on results from Type-II CT disk formation simulations by suggesting 
that U-shaped age profiles accompany the formation of \textit{all} manner of 
Type II breaks in disk galaxies, particularly those whose breaks are associated 
with either a bar or disk asymmetry (which drive the observed correlation).  
Whether all classes of Type II disks follow a unified r$_{\text{Min}}$-r$_{\text{Brk}}$ correlation is an issue to be explored further by models and observations 
aimed at a global understanding of the Type II phenomenon.  For instance, it 
will be of interest to explore why the observed correlation is not one-to-one 
(the best-fit slope differs from unity at the 2$\sigma$ level).  Does the fact 
that r$_{\text{Min}}$ falls either within or outside of r$_{\text{Brk}}$ for most of 
the eight galaxies plotted in \Fig{rMinvsrBrk} signal that some new mechanisms 
must be accounted for in future simulations?  In this sense, it is noteworthy 
that \cite{Yo12} find that two of their three U-shaped age profile detections 
have the minima occuring well inside of these galaxies' corresponding \textit{Type II-CT} breaks.  Conversely, the one Type II-CT from our sample which appears 
in \Fig{rMinvsrBrk} lies almost directly on top of the 1:1 lime, as found in 
simulations of this particular Type II class.

The existence of an r$_{\text{Min}}$-r$_{\text{Brk}}$ correlation amongst all Virgo 
Type II disks suggests that a common physical mechanism is responsible for both 
phenomena in any given Type II galaxy.  In the case of classical truncations, 
R08 argued that the break in their simulated Type II disk resulted from the 
combination of angular momentum conservation within the collapsing gas cloud, 
an assumed surface density threshold for star formation and resonant scattering 
of stars to radii beyond the truncation of the galaxy's star-forming disk.  
Since older stars would have more time to diffuse to greater distances than 
younger ones, R08 naturally found that the break in their galaxy coincided with 
an inversion in its age profile.  On the other hand, while very similar 
mechanisms were considered in SB09's simulation, these authors offered another 
explanation for the origin of their galaxy's break, whereby the rapid decline 
in star formation beyond the break radius (also seen in R08) was due to the 
onset of a warp in the gas disk rather than angular momentum conservation (as 
in R08).  Although radial migrations still deposited a significant number of 
stars beyond the break in SB09's simulation, these authors identified that the 
age minimum in their simulated galaxy was robust to whether radial migrations 
are invoked or not.  Unfortunately, we could not assess with our data the level 
at which radial migrations contribute to the r$_{\text{Min}}$-r$_{\text{Brk}}$ 
correlation reported here since each of the recent simulations of Type II disk 
formation predict U-shaped age profiles within them and have only addressed the 
case of classical truncations thus far.

While ours is not the first report of a connection between Type II breaks and U-shaped age profiles, we have found evidence to suggest that this connection 
extends to \textit{all} classes of Type IIs (not just Type II-CTs) through 
explicit age determinations.  To our knowledge, the only previous detections of 
U-shaped age profiles in Type IIs are from \cite{Yo10,Yo12}.  Other previous 
similar age gradient studies of disk galaxies either only treated one galaxy at 
a time \citep{dJ07,Wi09,Ba11,RS12} or were hampered by signal-to-noise 
constraints and uncertain age determinations (due to the age-metallicity 
degeneracy at optical wavelengths, which is lifted by use of near-infrared 
fluxes, as we have done here) \citep{Az08,Ba08}.

If the observed r$_{\text{Min}}$-r$_{\text{Brk}}$ correlation among Virgo Type IIs 
truly indicates that the formation of Type II breaks and U-shaped age profiles 
is causally connected, then the paucity of age minima reported in \Table{TypeIIDat} clearly demands an explanation (positive, flat and negative gradients 
describe the age profiles for 36\%, 14\% and 14\% of our Type II sub-sample, 
respectively).  This absence is most acutely noticed amongst the seven Type II-CTs in our sub-sample since U-shaped age profiles seem to be a robust prediction 
from the multiple simulations of these galaxies' formation performed thus far, 
yet we find such a profile in only one of these systems (VCC 0865).  Of course, 
our comparison could be flawed by the fact that those simulations all addressed 
Type II-CT formation within the field, while the environment in which our 
cluster galaxies reside likely plays a significant role in their evolution.  It 
is however worth recalling that in their stellar population analysis of \textit{field} Type IIs, \cite{Yo12} found that three of their six systems also lack U-shaped age profiles.  These two arguments then suggest that some mechanism(s) 
(perhaps environmentally-triggered; \eg harassment) either inhibit the 
formation of, or erase the age minima within, all Type II classes.  For 
instance, the age profiles for five of the eight Virgo Type IIs exhibiting 
positive age gradients (VCC 0522, 0945, 1126, 1726 and 2058) are fairly flat in 
their interiors but climb to relatively old ages in their outskirts; perhaps 
these galaxies' age profiles once more closely resembled that of VCC 1987 
(\Fig{APrfs-TypeII}).  Therefore, two fundamental issues for the future 
exploration of Type II formation models prevail: (i) whether U-shaped age 
profiles are expected in all Type II galaxy classes, and (ii) under what 
conditions are such age profiles prevented or simply erased?

\subsection{Type IIIs}\label{sec:R&D-TypeIIIs}
The mean stellar age profiles for the remaining disk galaxy type in our sample, 
the Type IIIs (of which we have 22), are shown in \Fig{APrfs-TypeIII}.  
Unfortunately, since Type III breaks typically occur at low surface 
brightnesses ($\sim$27-28 \magarc), our age profiles for these galaxies often 
fail to extend to, or much beyond, their break radii.  Still, over the radial 
extent probed by their age profiles, we find that (similar to the case of our 
Type II sub-sample) Virgo Type IIIs are described by a variety of age 
gradients: 23\% are flat, 32\% are positive, 18\% are negative and 27\% are U-shaped (\Table{AGrdFrctns}); the inner and outer gradients measured for those 
galaxies having U-shaped age profiles are listed in \Table{AgeGrads}, while the 
break and age inversion radii (where applicable) for our entire Type III sub-sample can be found in \Table{TypeIIIDat}.  Given that hydrodynamical simulations 
of Type III disk formation are scant (\citealt{Yo07}; and predictions of their 
stellar population properties are sorely lacking), the only possible 
extrapolation to our observations is that a variety of age gradients likely 
implies a corresponding variety of formation channels.  It will be of interest 
to see whether future Type III disk formation models can reproduce our observed 
age gradient distribution and if environmentally-triggered processes (\eg 
harassment) are needed to achieve such an agreement.

Arguably the most interesting result to come out of our inspection of the age 
profiles for Virgo Type IIIs is that a significant fraction (27\%) of them 
exhibit statistically-significant U-shapes, akin to those seen in formation 
models of Type II-CT disks.  In this case, however, we find that the minima 
systematically occur well inside these galaxies' breaks, as judged from the 
locations of the orange (breaks) and blue (minima) lines in 
\Fig{APrfs-TypeIII}.  This agrees with \cite{Ba08} who found that the stacked 
(optical) colour profile for Type IIIs in the field exhibits an inversion at 
roughly half the distance to their break radii.  As for our Type Is and IIs, we 
verified (through bootstrap linear fits) that each of proposed minima in our 
Type III sub-sample were defined by statistically-meaningful negative and 
positive age gradients leading into and out of them, respectively (\Table{AgeGrads}).

Since simulations of Type II-CT formation suggest that radial migrations foster 
U-shaped age profiles within these galaxies, we might speculate that a similar 
process is responsible for such profiles within Type III disks.  In fact, 
\cite{Yo07} have shown that a Type III luminosity profile could be produced 
through a minor merger onto a gas-rich disk, whereby the stars found beyond the 
remnant's break come predominantly from the progenitor disk.  Thus, the merger 
event effectively incites a migration of stars from the inner regions of the 
progenitor to the outer regions of the remnant.  Assuming the progenitor forms 
inside-out, this migratory process should raise the mean age of stars in the 
remnant's outskirts, thus giving rise to a U-shaped age profile.  A minor 
merger origin for Type III disks is further supported by the observation that 
five of the six Virgo Type IIIs exhibiting U-shaped age profiles also host 
bars, a phenomenon well known to be created, at least in part, by the 
pericenter passages of dwarf satellites.  Furthermore, dynamical simulations of 
barred galaxies have revealed that an age minimum should appear in the vicinity 
of the bar's ultra-harmonic and outer Lindblad resonances \citep{Wo07}.  The 
bars in these Virgo Type IIIs may then have played a key role in driving the 
radial migrations that we suspect were a principal component of their formation 
and might also explain why their age minima are found well into their interior 
regions.  Despite our best efforts though, understanding these observations 
would clearly benefit from more extensive Type III formation models.

\subsection{The role of environment in disk galaxy evolution}
\label{sec:R&D-Enviro}
The data-model comparisons carried out in each of the three previous sections 
all point to the common interpretation that the evolution of disk galaxies is 
likely very different in field and cluster environments.  While this idea is 
already well ingrained, our comparisons of the mean stellar age profiles for 
all types of Virgo disk galaxies against predictions from hydrodynamic 
simulations for field galaxies (and along with the fractions of each disk Type 
as a function of environment; \se{Data}) further bolster this claim.

Our argument can be extended with a plot in \Fig{DM87Distbn} of the 
differential distribution of (projected) cluster-centric radii for our entire 
Virgo disk galaxy sample, where galaxies have been binned and coloured 
according to their age gradients (flat, positive, negative, U-shaped).  The 
point types represent their luminosity profiles shape: Type I, II, III.  The 
coloured vertical lines correspond to the median value of each distribution.  
Despite global projection effects, \Fig{DM87Distbn} shows that Virgo cluster 
galaxies with flat and positive age gradients are located preferentially closer 
to the cluster center than those with negative and U-shaped gradients.  Given 
our expectations that, in the field, Type I disks would harbour negative age 
gradients and Type II and III disks would have U-shaped gradients, a simple 
overall explanation is that all of these galaxies initially formed like their 
field counterparts, but upon entering the cluster, their (local) environments 
transformed their age gradients into flat and/or positive ones.  Conceptually 
speaking, such a transformation could be induced by ram pressure stripping and 
harassment; processes which would conceivably quench the star formation in 
these galaxies' outer disks (thereby raising the mean stellar age there) and 
radially mix the stellar populations throughout their disks (thereby flattening 
any pre-existing gradients).  In order to better understand the observations 
presented here though, environmentally-driven processes must be considered in 
future simulations of disk galaxy formation.  Moreover, the observational case 
for a cluster-centric radial dependence of age gradients in disk galaxies could 
be improved (or disproved as the case may be) with deep, multi-band imaging for 
a larger sample of cluster galaxies with more accurate distances \citep{Fe12}.


\section{CONCLUSIONS}\label{sec:Conc}

We have analysed the luminosity and mean stellar age profiles for a complete, 
magnitude-limited sample of disk galaxies drawn from the Virgo cluster and 
shown that each of the three major disk galaxy types in this cluster can 
harbour statistically-meaningful inversions (``U-shapes'') in their age 
profiles, akin to those predicted by recent high-resolution simulations of Type 
II disk galaxy formation in the field \citep{Ro08,MS09,SB09}.  Since the 
inversions found in the simulated galaxies were formed, in part, by radial 
migrations of disk stars (and lacking guidance from simulations of cluster 
galaxies), we speculate that stellar migrations are a generic feature of galaxy 
formation, independent of both the disk galaxy type and the particular agent by 
which migrations are achieved (\ie bar, spiral arm, satellite).  However, the 
fact that all of our cluster disk galaxiess, most notably six of our seven classically-truncated Type IIs (as well as disks found in \textit{field} 
environments; \citealt{Yo12}), do not exhibit inversions in their age profiles 
and given the reduced fraction of Type II systems in the Virgo cluster relative 
to the field, suggests that the significance of stellar migrations is likely 
sensitive to other prominent factors involved in galaxy evolution, such as 
environmentally-driven processes in clusters (\eg harassment).  Equally 
interesting is that, in the absence of U-shaped age profiles, most of our 
galaxies are described by age profiles having either flat or positive 
gradients, as opposed to the negative gradients generically predicted by 
formation models of exponential disks.  We still interpret this observation as 
supporting a picture of disk galaxy evolution where environment plays a 
principal role.

We have also shown that, for those Type II cluster galaxies in our sample which 
exhibit significant inversions in their age profiles, the locations of their 
luminosity profile breaks and age minima are correlated.  This finding 
tentatively confirms the predictions of simulations of classically-truncated 
Type IIs \citep{Ro08,MS09,SB09} and sets the stage for further simulation work 
since the above correlation suggests that process(es) causing \textit{all} 
manners of Type II breaks in disk galaxies would also generate inverted age 
profiles.  However, while the simulations imply a direct (1:1) correspondence 
between the light profile break and age inversion radii, we find a 
significantly steeper slope ($1.6\pm0.3$) for those two parameters.  This 
result may indicate additional physics at play in the age profile inversion 
phenomenon or simply reflect some uncertainty in our (visual) assessment of 
those breaks and/or minima.  Still, although our hypothesis for a \textit{unified} ``light profile break-age profile inversion'' correlation amongst all Type 
II disks is new, it complements previous findings of age inversions in 
\textit{classically-truncated} Type IIs well inside of their respective breaks.

Our results on the stellar populations of all three disk galaxy types within 
the cluster environment pose a challenge to current galaxy formation models, 
which we are likely explained by environmentally-induced evolutionary processes 
still missing in such models.  In particular, future models may possibly 
reproduce the distribution of disk galaxy types and the respective mixture of 
age gradients in cluster galaxies once effects like ram pressure stripping and 
harassment have been accounted for.

In an Appendix, we concur with \cite{Er12} that Type II breaks are seemingly 
absent in cluster S0 galaxies.  We also view many of their Type III 
classifications as either single $n>1$ S\'ersic profiles, bona fide disk 
upturns or Type I disk systems with a faint stellar halo.  The latter ought to 
be routinely detected in galaxy images reaching below $\mu_i \simeq 27$ 
\magarc, as in M31 \citep{Co11}.

\bigskip

We thank Peter Erwin, Kelly Foyle, Brad Gibson and Robert Thacker for comments 
on an earlier version of this paper which led to valuable improvements.  The 
referee's thoughtful suggestions also improved the content and clarity of our 
paper.  J.R. and S.C. acknowledge financial support from the National Science 
and Engineering Council of Canada through a postgraduate scholarship and a 
Discovery Grant, respectively.  P.S.B. is supported by the Ministerio de 
Ciencia e Innovacion (MICINN) of Spain through the Ramon y Cajal programme.  
P.S.B. also acknowledges a Marie Curie Intra-European Reintegration grant 
within the 6th European framework program.  Support for this work was provided 
to M.M. by NASA through SAO Award Number 2834-MIT-SAO-4018, which is issued by 
the Chandra X-ray Observatory on behalf of NASA under contract NAS8-03060.


\section{APPENDIX: \cite{Er12}}\label{App:Er12}

We noted in \se{Data} that the 55 S0 galaxies from the \cite{MD11} sample were 
left out of our analysis.  This is due to the complicated interpretation of 
their surface brightness profiles where the bulge component is typically 
dominant.  According to the NIR bulge-disk decompositions of \cite{MD09}, 
nearly two-thirds of the 55 S0s in our parent sample exhibit exponential-like 
profiles (within their sky error envelopes).  Thus, without additional chemical 
or dynamical information, these structures could either be viewed as bulge- or 
disk-like.  Given the inherent challenges with analyzing S0's faint outer disks 
and their acute sensitivity to sky levels, accurate modelling of their bulges 
via multi-band bulge-disk decompositions would be required in the pursuit of 
such a goal, a level of detail which lies beyond the scope of the present work 
(see also \citealt{PC11}). 

Given the non-trivial interpretation of the outer disks in S0 galaxies, it 
seems prudent to examine the recent claims by Er12 about Virgo cluster S0 
disks.  Using optical imaging largely from the Sloan Digital Sky Survey, as 
well as some of their own observations, Er12 compared the incidence of Type II 
breaks in field and cluster S0s to find that such features are entirely absent 
in the cluster S0s (at least down to a limiting surface brightness of $r \sim$ 
27 \magarc).  Rather, these authors find that the luminosity profiles of 
cluster S0s are evenly distributed between the Type I and III classes.  \Fig{SBPrfs-Er12} shows the $griH$ surface brightness profiles for the 18 S0 galaxies 
from our parent sample which overlap with Er12's Virgo sample.  Our 
classification of pure exponential (Type I) luminosity profiles for VCC 654, 
778, 784, 1253, 1720, 1813 and 2092 agrees with Er12.  Other luminosity profile 
shapes, such as Er12's Type IIIs, are however less clearly agreed upon.  For 
instance, VCC 523, 1154, 1196 and 1537 have no discernible profile break or 
outer disk component and are modeled as a single \Sersic structure in 
\cite{MD11}; Er12 instead view them as Type III systems (with breaks located at 
27\arcsec, 119\arcsec, 63\arcsec and 29\arcsec).  Er12 also identify Type III 
profiles (breaks) in VCC 355 (83\arcsec), 759 (190\arcsec), 1827 (54\arcsec) 
and 1902 (58\arcsec), yet our sky error envelopes, as shown in 
\Fig{SBPrfs-Er12}, make them consistent with both a single exponential or a 
Type III disk.  An issue with our profile shape assessments is the relatively 
low signal-to-noise of the SDSS images at levels where Type III breaks are 
normally detected.

Whether galaxies with outer profile upturns are viewed as a smooth \Sersic 
bulge plus a Type III disk or a \Sersic bulge plus a Type I disk and a stellar 
halo depends highly on the nature of the isophotal fitting in the inner parts 
(\eg whether isophotes have concentric [Erwin] or variable [us] position 
angles), the accuracy of sky subtraction, and the ability to distinguish a 
stellar disk from a stellar halo.

This is not to say that disk profile upturns, or ``anti-truncated'' disks, are 
unexpected.  These may result from a component which is actually independent 
from the disk itself, such as a stellar halo (\eg as seen in M31; 
\citealt{Co11}) or other accreted external material.  Such profile upturns, or 
disk/halo transitions, are typically seen at surface brightnesses below $\mu_i 
\simeq 26-27$ \magarc\ \citep{Co11}.  With a typical limit of $\mu_i \simeq 
26.5$ \magarc\ \citep{Ha12}, our own SDSS profiles are barely deep enough to 
assess the presence of disk/halo structures.  The latter is clearly of 
tremendous interest, though once again, beyond the scope of the present study.

Type III upturns may also be purely confined to the disk, as confirmed by 
spiral-arm structure which appears to extend beyond the Type III break radius 
(\eg NGC 3982 in \citealt{Er08}; see their Fig. 11).  Models which explain Type 
III disks via mechanisms other than pure accretion include harassment or minor 
mergers.  The harassment simulations of \cite{Mo99} show clearly the evolution 
of a Type I disk into a Type III system.   The minor merger models of 
\cite{Yo07} also show that the ``outer light excess'' comes from disk stars in 
the original primary galaxy that gained angular momentum from the minor merger 
process.  All things considered, though, the generation of a Type III profile 
always requires an external agent.

In summary, while we do not find Type III systems in our Virgo cluster sample 
of S0 galaxies \citep{MD11}, we concur with Er12's important claim that Type II 
breaks are seemingly absent in S0 cluster galaxies.  The latter is confirmed by 
examination of our Virgo cluster sample \citep{MD11}.


\clearpage


\begin{deluxetable}{ccccc}
 \tabletypesize{\scriptsize}
 \tablewidth{0pc}
 \tablecaption{Luminosity profile types as a function of morphology for Virgo disk galaxies}
  \tablehead{
  \colhead{Type} &
  \multicolumn{3}{c}{Fraction per morphological bin (\%)} &
  \colhead{Total} \\
  \colhead{} &
  \colhead{Sa$-$Sbc} &
  \colhead{Sc$-$Sm} &
  \colhead{Im} &
  \colhead{fraction (\%)} \\
  \colhead{(1)} &
  \colhead{(2)} &
  \colhead{(3)} &
  \colhead{(4)} &
  \colhead{(5)}
 }
 \startdata
   I & 0.03 & 0.14 & 0.14 & 0.31 \cr
  II & 0.09 & 0.14 & 0.11 & 0.34 \cr
 III & 0.19 & 0.12 & 0.03 & 0.34 \cr
 \enddata
 \label{tbl:TypeFrctns}
\end{deluxetable}

\begin{deluxetable}{ccc}
 \tabletypesize{\scriptsize}
 \tablewidth{0pc}
 \tablecaption{Inner and outer mean stellar age gradients for Virgo disk galaxies hosting U-shaped age profiles.}
 \tablehead{
  \colhead{VCC} &
  \colhead{d$<$A$>$/d$r_{\text{i}}$} &
  \colhead{d$<$A$>$/d$r_{\text{o}}$} \\
  \colhead{ID} &
  \colhead{(Gyr kpc$^{-1}$)} &
  \colhead{(Gyr kpc$^{-1}$)} \\
  \colhead{(1)} &
  \colhead{(2)} &
  \colhead{(3)}
 }
 \startdata
 \multicolumn{3}{c}{Type I} \cr
 0583 & -5.70            & +1.78 $\pm$ 0.51 \cr
 1410 & -2.13 $\pm$ 0.40 & +3.05 $\pm$ 0.38 \cr
 1566 & -6.88 $\pm$ 1.80 & +0.56 $\pm$ 0.30 \cr
 \multicolumn{3}{c}{Type II} \cr
 0596 & -0.32 $\pm$ 0.02 & +0.05 $\pm$ 0.02 \cr
 0849 & -3.61 $\pm$ 0.72 & +1.25 $\pm$ 0.42 \cr
 0865 & -0.19 $\pm$ 0.04 & +0.56 $\pm$ 0.02 \cr
 1021 & -2.04            & +1.29 $\pm$ 0.29 \cr
 1654 & -1.09            & +4.72 $\pm$ 0.95 \cr
 1929 & -0.58 $\pm$ 0.18 & +0.54 $\pm$ 0.07 \cr
 1943 & -0.59 $\pm$ 0.16 & +0.35 $\pm$ 0.11 \cr
 1987 & -0.40 $\pm$ 0.03 & +0.48 $\pm$ 0.02 \cr
 \multicolumn{3}{c}{Type III} \cr
 0692 & -2.04 $\pm$ 0.78 & +0.88 $\pm$ 0.30 \cr
 0912 & -0.78 $\pm$ 0.34 & +0.31 $\pm$ 0.04 \cr
 1379 & -0.60 $\pm$ 0.11 & +1.10 $\pm$ 0.11 \cr
 1686 & -1.82 $\pm$ 0.37 & +2.26 $\pm$ 1.18 \cr
 1811 & -1.67 $\pm$ 0.46 & +0.54 $\pm$ 0.23 \cr
 2012 & -0.28 $\pm$ 0.08 & +0.42 $\pm$ 0.35 \cr
 \enddata
 \label{tbl:AgeGrads}
\end{deluxetable}

\begin{deluxetable}{ccrrr}
 \tabletypesize{\scriptsize}
 \tablewidth{0pc}
 \tablecaption{Morphologies, effective radii and age minimum radii (when present) for Virgo Type I galaxies.}
 \tablehead{
  \colhead{VCC} &
  \colhead{Hubble} &
  \colhead{r$_{\text{e}}$} &
  \multicolumn{2}{c}{r$_{\text{Min}}$} \\
  \cline{4-5}
  \colhead{ID} &
  \colhead{type} &
  \colhead{(kpc)} &
  \colhead{(kpc)} &
  \colhead{($r_{\text{e}}$)} \\
  \colhead{(1)} &
  \colhead{(2)} &
  \colhead{(3)} &
  \colhead{(4)} &
  \colhead{(5)}
 }
 \startdata
 0510 & Sd         & 1.495 &   ... &   ... \cr
 0583 & Im         & 1.594 & 1.795 & 1.126 \cr
 0825 & Im         & 1.492 &   ... &   ... \cr
 0857 & SB(r)b     & 3.209 &   ... &   ... \cr
 0980 & Im         & 2.272 &   ... &   ... \cr
 1047 & SB(rs)a    & 1.831 &   ... &   ... \cr
 1410 & Scd        & 0.985 & 1.469 & 1.492 \cr
 1448 & Im         & 3.120 &   ... &   ... \cr
 1450 & IB(s)m     & 2.647 &   ... &   ... \cr
 1524 & SAB(s)m    & 3.179 &   ... &   ... \cr
 1566 & Sdm        & 2.510 & 1.257 & 0.501 \cr
 1569 & Scd        & 1.320 &   ... &   ... \cr
 1585 & IBm        & 1.645 &   ... &   ... \cr
 1588 & SAB(rs)cd  & 2.619 &   ... &   ... \cr
 1678 & Sm         & 2.412 &   ... &   ... \cr
 1684 & Scd        & 1.615 &   ... &   ... \cr
 1825 & Im         & 0.860 &   ... &   ... \cr
 1890 & Im         & 1.673 &   ... &   ... \cr
 1897 & Sd         & 2.157 &   ... &   ... \cr
 1931 & Im         & 1.530 &   ... &   ... \cr
 \enddata
 \label{tbl:TypeIDat}
\end{deluxetable}

\begin{deluxetable}{cccrrrrr}
 \tabletypesize{\scriptsize}
 \tablewidth{0pc}
 \tablecaption{Morphologies, effective radii, break radii and age minimum radii (when present) for Virgo Type II galaxies.}
 \tablehead{
  \colhead{VCC} &
  \colhead{Hubble} &
  \colhead{Profile} &
  \colhead{r$_{\text{e}}$} &
  \multicolumn{2}{c}{r$_{Brk}$} &
  \multicolumn{2}{c}{r$_{Min}$} \\
  \cline{5-6}
  \cline{7-8}
  \colhead{ID} &
  \colhead{type} &
  \colhead{type$^{\text{1}}$} &
  \colhead{(kpc)} &
  \colhead{(kpc)} &
  \colhead{($r_{\text{e}}$)} &
  \colhead{(kpc)} &
  \colhead{($r_{\text{e}}$)} \\
  \colhead{(1)} &
  \colhead{(2)} &
  \colhead{(3)} &
  \colhead{(4)} &
  \colhead{(5)} &
  \colhead{(6)} &
  \colhead{(7)} &
  \colhead{(8)}
 }
 \startdata
 0522 & SAB(r)a    & o-OLR+III &  2.396 &  3.297 & 1.376 &   ...  &  ...  \cr
 0596 & SAB(s)bc   & o-AB+III  &  8.398 &  8.242 & 0.981 &  8.554 & 1.019 \cr
 0620 & Sm         & AB        &  1.157 &  0.577 & 0.497 &   ...  &  ...  \cr
 0664 & IB(s)m     & o-CT      &  2.014 &  5.357 & 2.660 &   ...  &  ...  \cr
 0792 & S(rs)b     & CT+III    &  4.019 &  5.769 & 1.434 &   ...  &  ...  \cr
 0849 & SAB(s)c    & AB+III    &  1.720 &  2.102 & 1.222 &  1.790 & 1.041 \cr
 0865 & Sd         & CT+III    &  6.716 &  5.522 & 0.822 &  5.188 & 0.773 \cr
 0945 & Im         & AB        &  1.947 &  1.154 & 0.592 &   ...  &  ...  \cr
 1021 & IABm       & i+III     &  1.273 &  0.824 & 0.604 &  0.587 & 0.428 \cr
 1126 & Sb         & CT+III    &  2.280 &  3.297 & 1.445 &   ...  &  ...  \cr
 1200 & Im         & AB        &  1.412 &  0.742 & 0.437 &   ...  &  ...  \cr
 1508 & SB(rs)d    & o-CT+III  &  2.486 &  5.769 & 2.321 &   ...  &  ...  \cr
 1532 & SB(rs)c    & o-AB+III  &  2.510 &  1.648 & 0.657 &   ...  &  ...  \cr
 1555 & SAB(s)c    & o-AB      & 20.877 & 13.187 & 0.632 &   ...  &  ...  \cr
 1654 & IABm       & o-AB      &  1.110 &  0.742 & 0.666 &  0.571 & 0.513 \cr
 1696 & S(r)d      & CT+III    &  3.043 &  6.181 & 2.031 &   ...  &  ...  \cr
 1726 & Im         & AB        &  1.738 &  1.484 & 0.838 &   ...  &  ...  \cr
 1929 & SAB(s)dm   & o-OLR+III &  2.871 &  2.225 & 0.775 &  1.664 & 0.580 \cr
 1943 & SAB(rs)bc  & i+III     &  2.151 &  2.225 & 1.034 &  2.741 & 1.276 \cr
 1952 & Im         & AB        &  0.959 &  1.154 & 1.204 &   ...  &  ...  \cr
 1987 & SAB(rs)cd  & o-AB+III  &  4.940 &  6.593 & 1.335 &  8.554 & 1.732 \cr
 2058 & S(rs)bc    & CT+III    &  4.026 &  4.533 & 1.126 &   ...  &  ...  \cr
 \enddata
 \tablenotetext{1}{i = break located within bar, o = break located outside bar, CT = classical truncation, AB = apparent/asymmetric break, OLR = break coincident with outer Lindblad resonance}
 \label{tbl:TypeIIDat}
\end{deluxetable}

\begin{deluxetable}{ccrrrrr}
 \tabletypesize{\scriptsize}
 \tablewidth{0pc}
 \tablecaption{Morphologies, effective radii, break radii and age minimum radii (when present) for Virgo Type III galaxies.}
 \tablehead{
  \colhead{VCC} &
  \colhead{Hubble} &
  \colhead{r$_{\text{e}}$} &
  \multicolumn{2}{c}{r$_{Brk}$} &
  \multicolumn{2}{c}{r$_{Min}$} \\
  \cline{4-5}
  \cline{6-7}
  \colhead{ID} &
  \colhead{type} &
  \colhead{(kpc)} &
  \colhead{(kpc)} &
  \colhead{($r_{\text{e}}$)} &
  \colhead{(kpc)} &
  \colhead{($r_{\text{e}}$)} \\
  \colhead{(1)} &
  \colhead{(2)} &
  \colhead{(3)} &
  \colhead{(4)} &
  \colhead{(5)} &
  \colhead{(6)} &
  \colhead{(7)}
 }
 \startdata
 0483 & S(rs)c     & 3.551 &  9.885 & 2.784 & 4.176 & 1.176 \cr
 0692 & SB(rs)ab   & 2.372 &  4.943 & 2.083 & 2.984 & 1.258 \cr
 0809 & Sdm        & 2.252 &  5.769 & 2.562 &   ... &   ... \cr
 0912 & SB(rs)ab   & 2.461 & 12.356 & 5.022 & 1.942 & 0.789 \cr
 1110 & S(s)ab     & 3.955 & 20.605 & 5.209 &   ... &   ... \cr
 1326 & SB(s)a     & 2.084 &  4.531 & 2.174 &   ... &   ... \cr
 1358 & Sa         & 1.188 &  3.709 & 3.121 &   ... &   ... \cr
 1379 & SAB(s)d    & 3.260 &  8.238 & 2.527 & 3.818 & 1.171 \cr
 1393 & SBcd       & 1.742 &  4.121 & 2.365 &   ... &   ... \cr
 1479 & SB(rs)ab   & 1.839 &  4.119 & 2.240 &   ... &   ... \cr
 1486 & SAB(s)b    & 0.900 &  4.121 & 4.579 &   ... &   ... \cr
 1575 & IBm        & 2.143 &  2.883 & 1.346 & 0.568 & 0.265 \cr
 1615 & SB(rs)b    & 5.427 & 16.484 & 3.037 &   ... &   ... \cr
 1686 & IBm        & 2.750 &  9.061 & 3.295 & 1.485 & 0.540 \cr
 1727 & SAB(rs)b   & 4.563 & 18.123 & 3.972 &   ... &   ... \cr
 1757 & SAB(s)a    & 1.614 &  5.766 & 3.573 &   ... &   ... \cr
 1811 & SAB(rs)b   & 1.562 &  4.119 & 2.637 & 2.188 & 1.401 \cr
 1972 & SAB(rs)c   & 7.806 &  6.181 & 0.792 &   ... &   ... \cr
 2012 & Scd        & 2.416 &  5.766 & 2.387 & 3.214 & 1.330 \cr
 2023 & Sc         & 2.663 &  4.533 & 1.702 &   ... &   ... \cr
 2042 & SABd       & 1.930 &  2.059 & 1.067 &   ... &   ... \cr
 2070 & S(s)ab     & 2.742 &  9.066 & 3.306 &   ... &   ... \cr
 \enddata
 \label{tbl:TypeIIIDat}
\end{deluxetable}

\begin{deluxetable}{ccccc}
 \tabletypesize{\scriptsize}  
 \tablewidth{0pc}  
 \tablecaption{Age gradient distributions for Virgo disk galaxies.}
 \tablehead{
  \colhead{Type} &
  \multicolumn{4}{c}{Age gradient fraction (\%)} \\
  \colhead{} &
  \colhead{Flat} &
  \colhead{Positive} &
  \colhead{Negative} &
  \colhead{U-shaped} \\
  \colhead{(1)} &
  \colhead{(2)} &
  \colhead{(3)} &
  \colhead{(4)} &
  \colhead{(5)}
 }
 \startdata
  I & 0.40 & 0.35 & 0.10 & 0.15 \cr
 II & 0.14 & 0.36 & 0.14 & 0.36 \cr
III & 0.23 & 0.32 & 0.09 & 0.36 \cr
 \enddata
 \label{tbl:AGrdFrctns}
\end{deluxetable}


\clearpage
\begin{figure*}
 \begin{center}
  \includegraphics[width=0.9\textwidth]{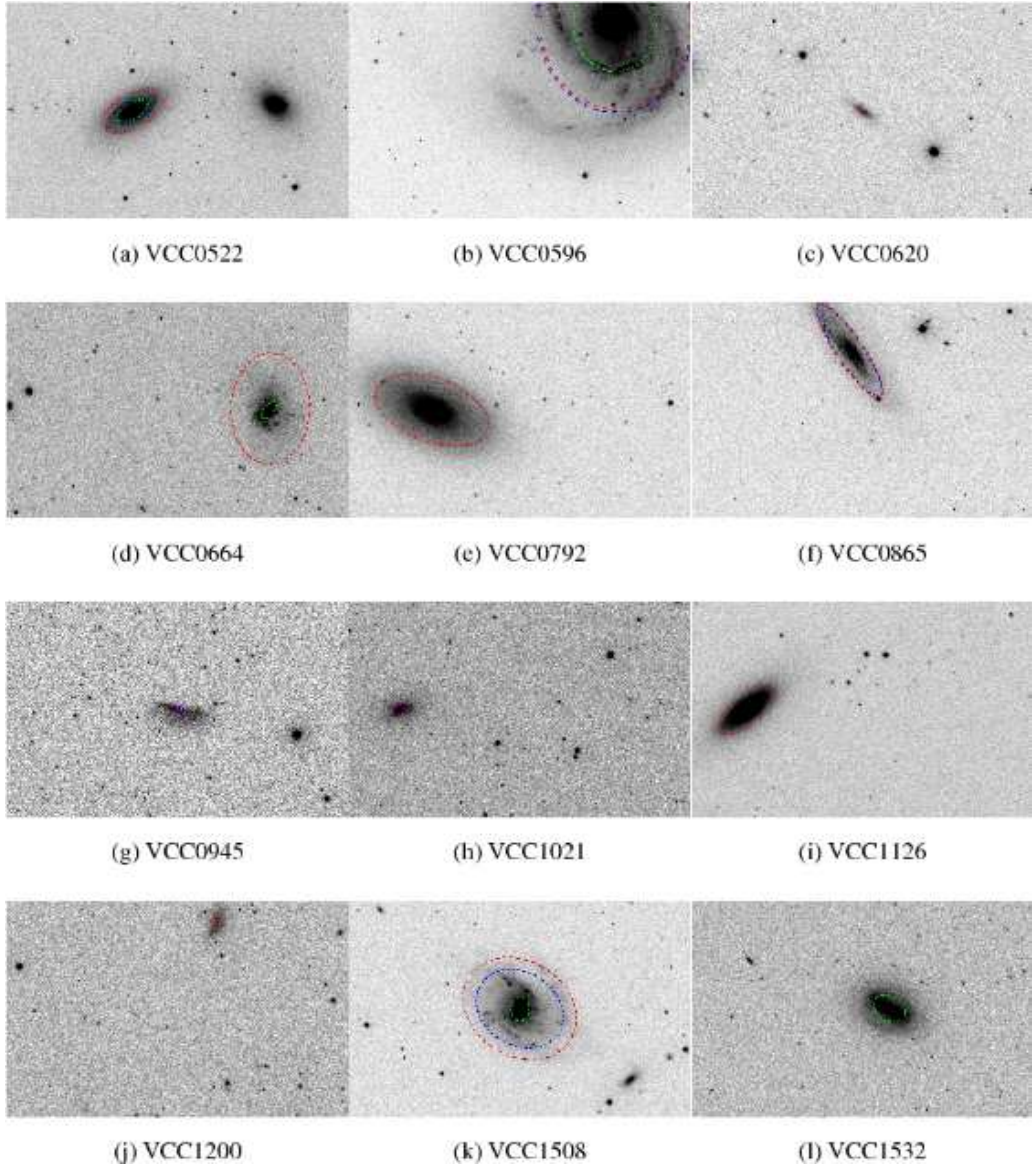}
  \caption{$i$-band images for our sub-sample of Virgo Type II galaxies.  The 
overlaid ellipses correspond to either the locations of the luminosity profile 
breaks (red) or mean stellar age profile minimum (if present; blue), or the 
extents of the bars (if present; green) within these systems.  The images are 
shown on a common pixel scale and have the same dimensions.}
  \label{fig:Images-TypeII}
 \end{center}
\end{figure*}

\clearpage
\begin{figure*}
 \ContinuedFloat
 \begin{center}
  \includegraphics[width=0.9\textwidth]{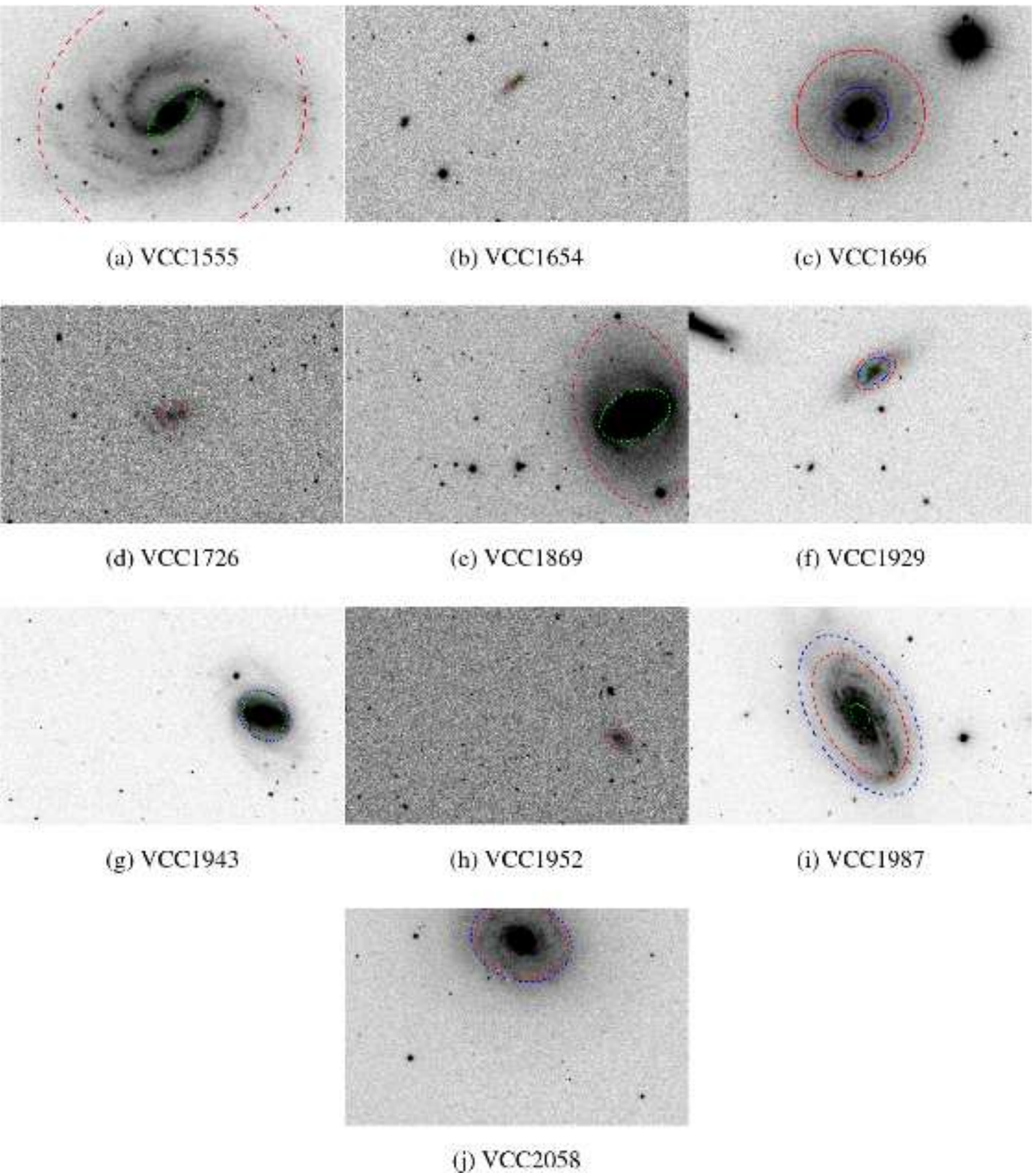}
  \caption{(\textit{continued})}
 \end{center}
\end{figure*}

\clearpage
\begin{figure*}
 \begin{center}
  \includegraphics[width=0.9\textwidth]{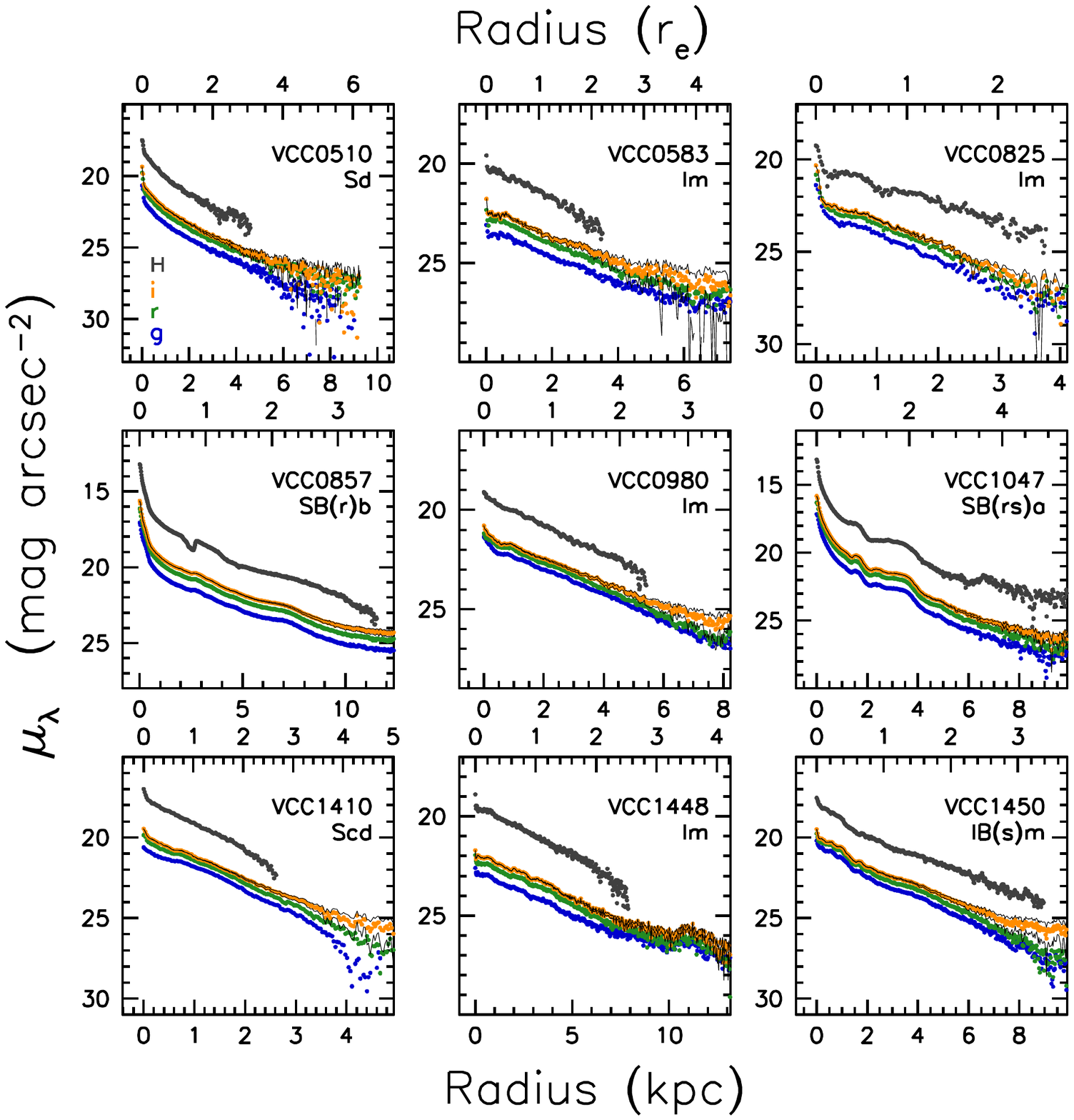}
  \caption{Surface brightness profiles in the $griH$ bands for our sub-sample 
of Virgo Type I galaxies ($g$ = blue; $r$ = green; $i$ = orange; $H$ = black).  
The Virgo Cluster Catalog (VCC; \citealt{Bi85}) identification number and 
detailed morphology is provided for each galaxy, while radii are quoted in 
terms of $H$-band effective radii \citep{MD09} and kiloparsecs on the top and 
bottom of each panel, respectively.  The 1$\sigma$ sky error envelope for each 
galaxy's $i$-band profile is shown as the black continuous line in each panel.}
  \label{fig:SBPrfs-TypeI}
 \end{center}
\end{figure*}

\clearpage
\begin{figure*}
 \ContinuedFloat
 \begin{center}
  \includegraphics[width=0.9\textwidth]{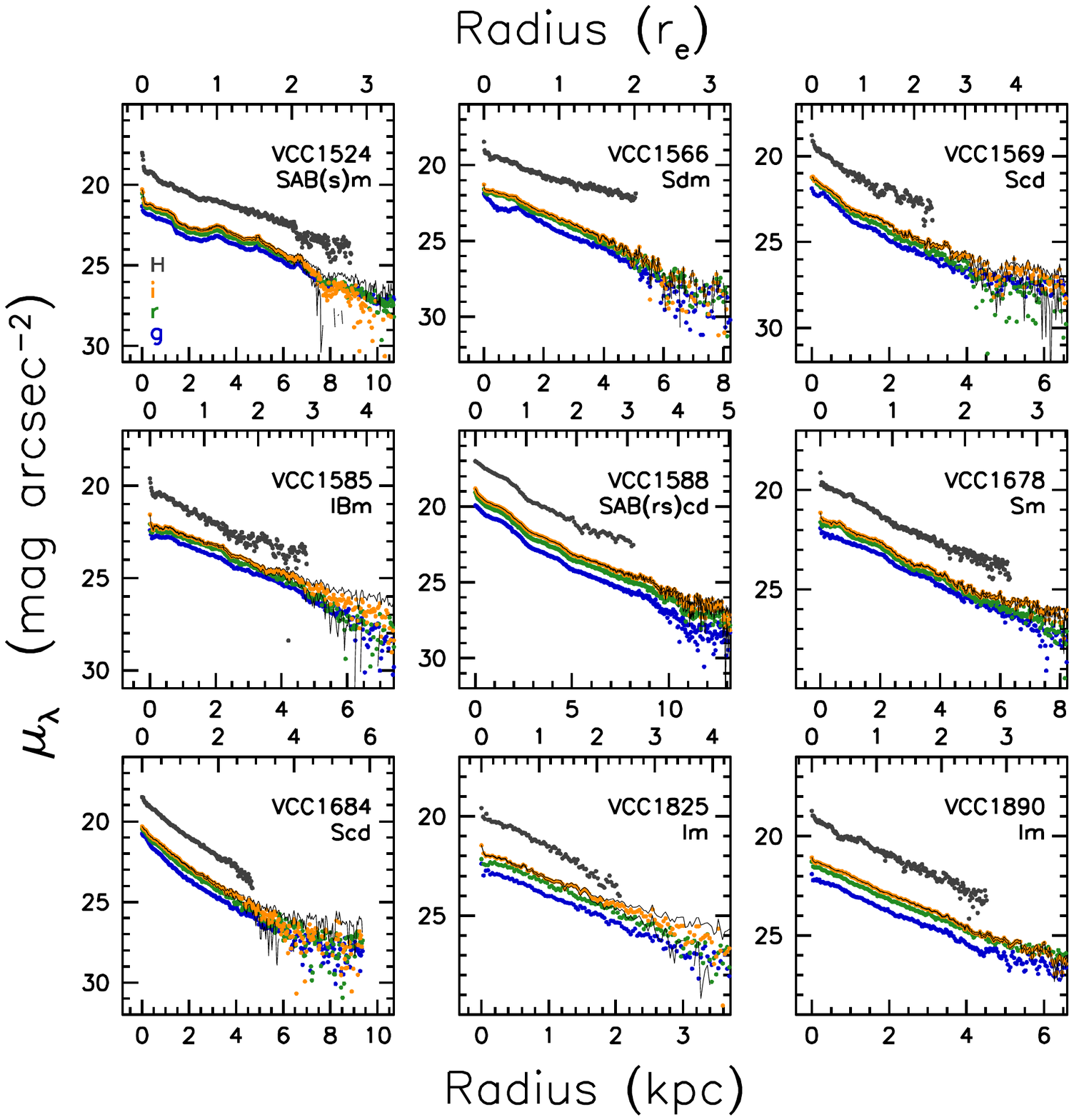}
  \caption{(\textit{continued})}
 \end{center}
\end{figure*}

\clearpage
\begin{figure*}
 \ContinuedFloat
 \begin{center}
  \includegraphics[width=0.9\textwidth]{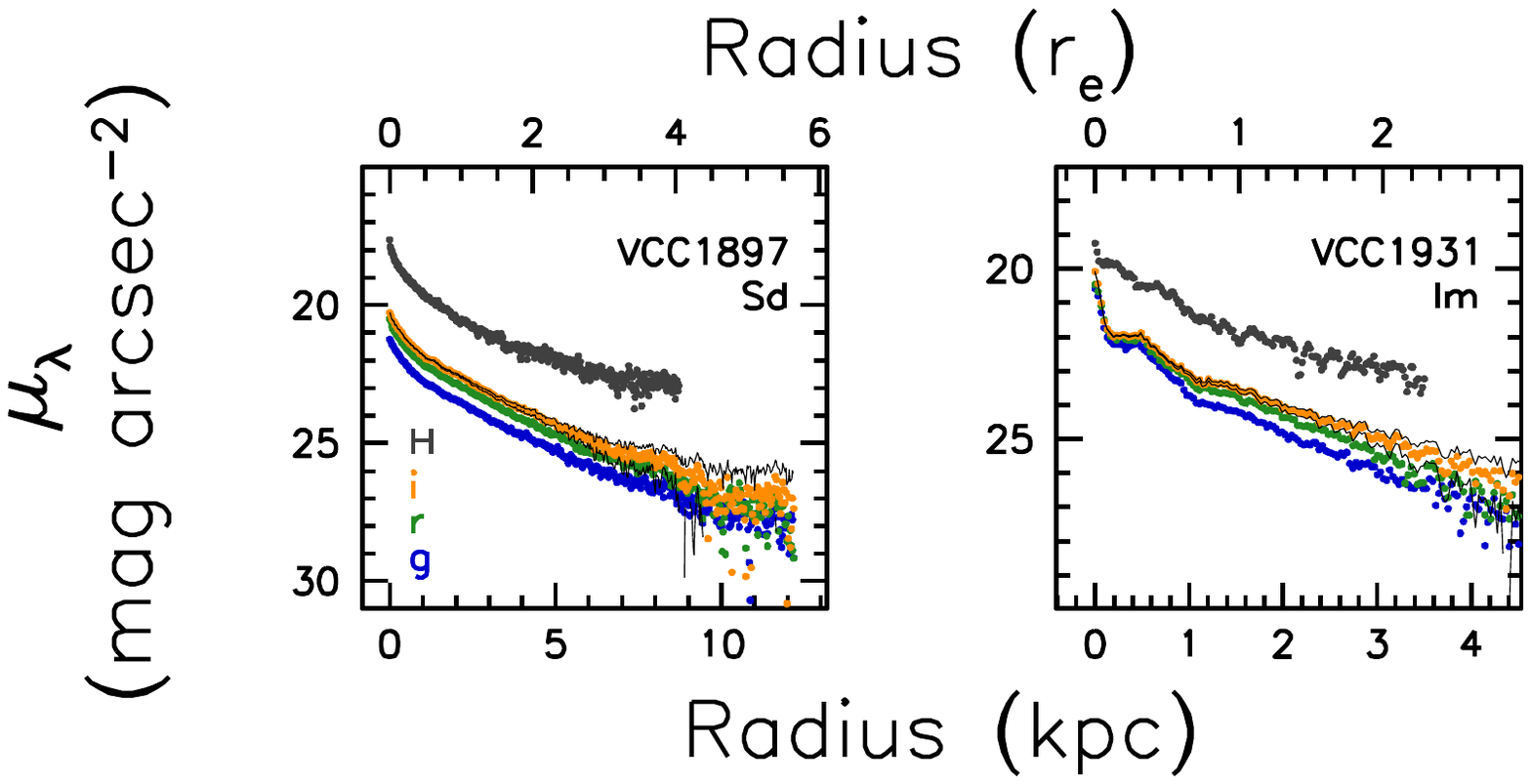}
  \caption{(\textit{continued})}
 \end{center}
\end{figure*}

\clearpage
\begin{figure*}
 \begin{center}
  \includegraphics[width=0.9\textwidth]{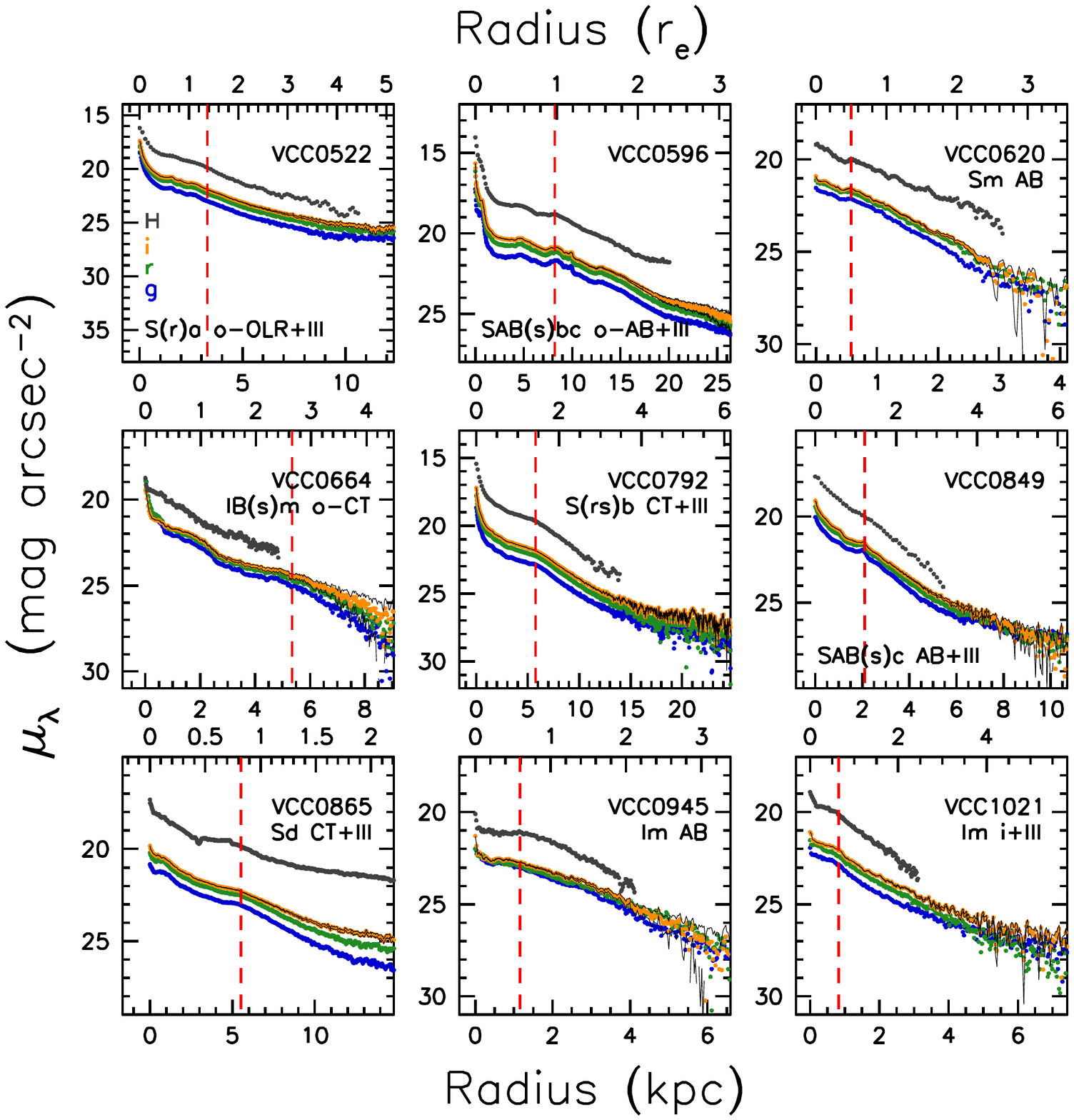}
  \caption{As in \Fig{SBPrfs-TypeI} but for Virgo Type II galaxies.  The 
specific class of Type II break has also been provided, while the location of 
the break for each galaxy is marked by the red vertical line.}
  \label{fig:SBPrfs-TypeII}
 \end{center}
\end{figure*}

\clearpage
\begin{figure*}
 \ContinuedFloat
 \begin{center}
  \includegraphics[width=0.9\textwidth]{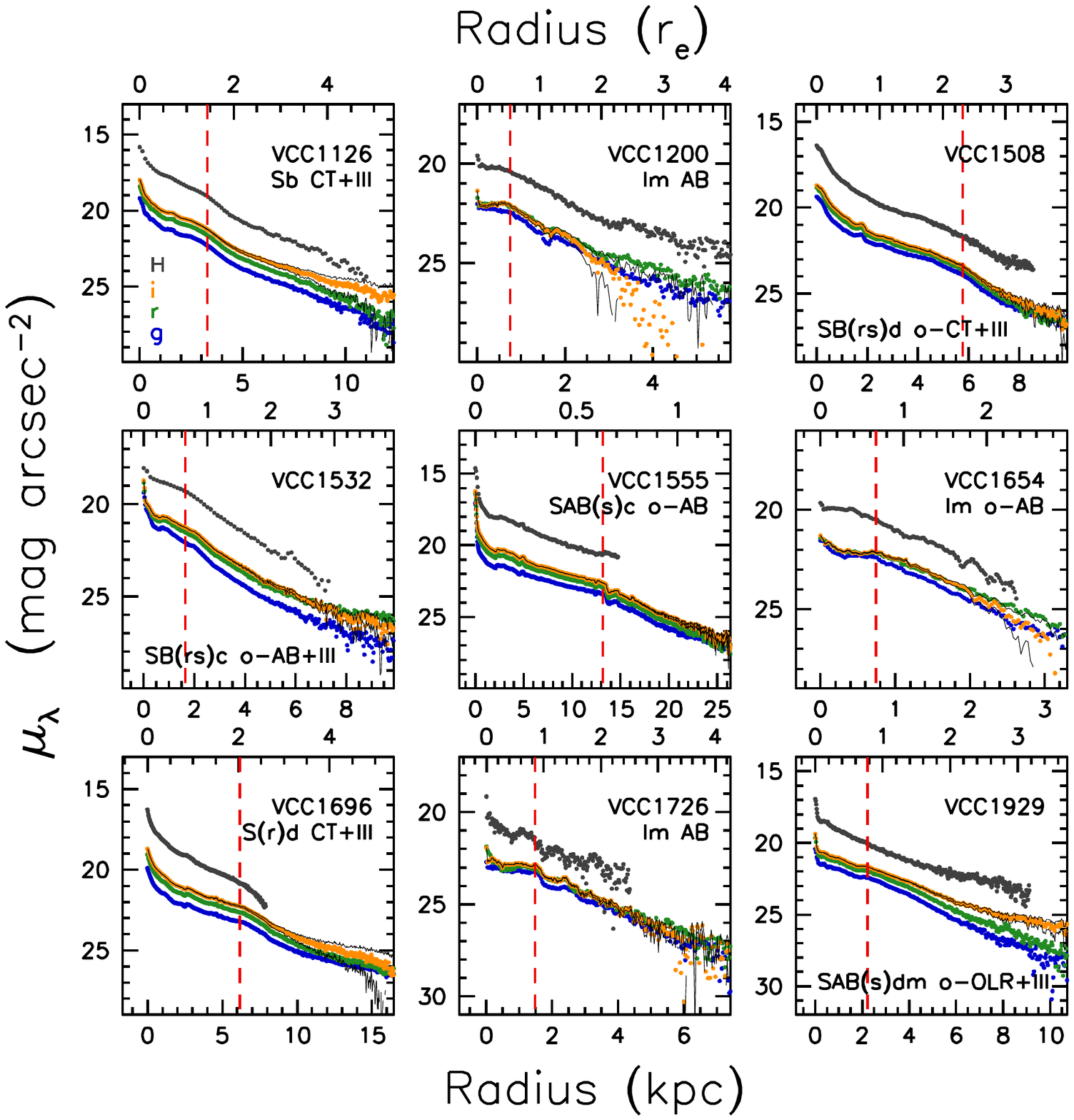}
  \caption{(\textit{continued})}
 \end{center}
\end{figure*}

\clearpage
\begin{figure*}
 \ContinuedFloat
 \begin{center}
  \includegraphics[width=0.9\textwidth]{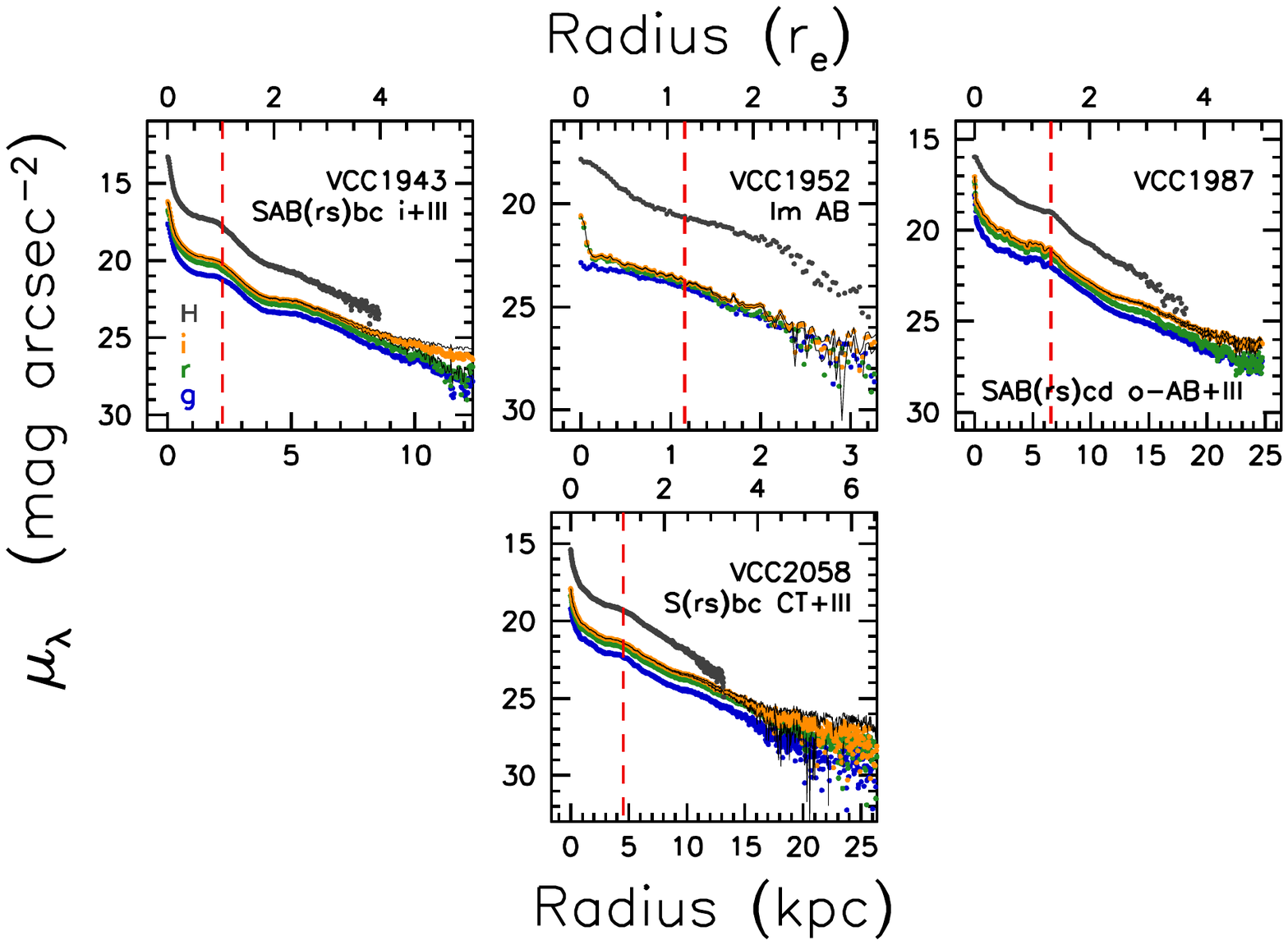}
  \caption{(\textit{continued})}
 \end{center}
\end{figure*}

\clearpage
\begin{figure*}
 \begin{center}
  \includegraphics[width=0.9\textwidth]{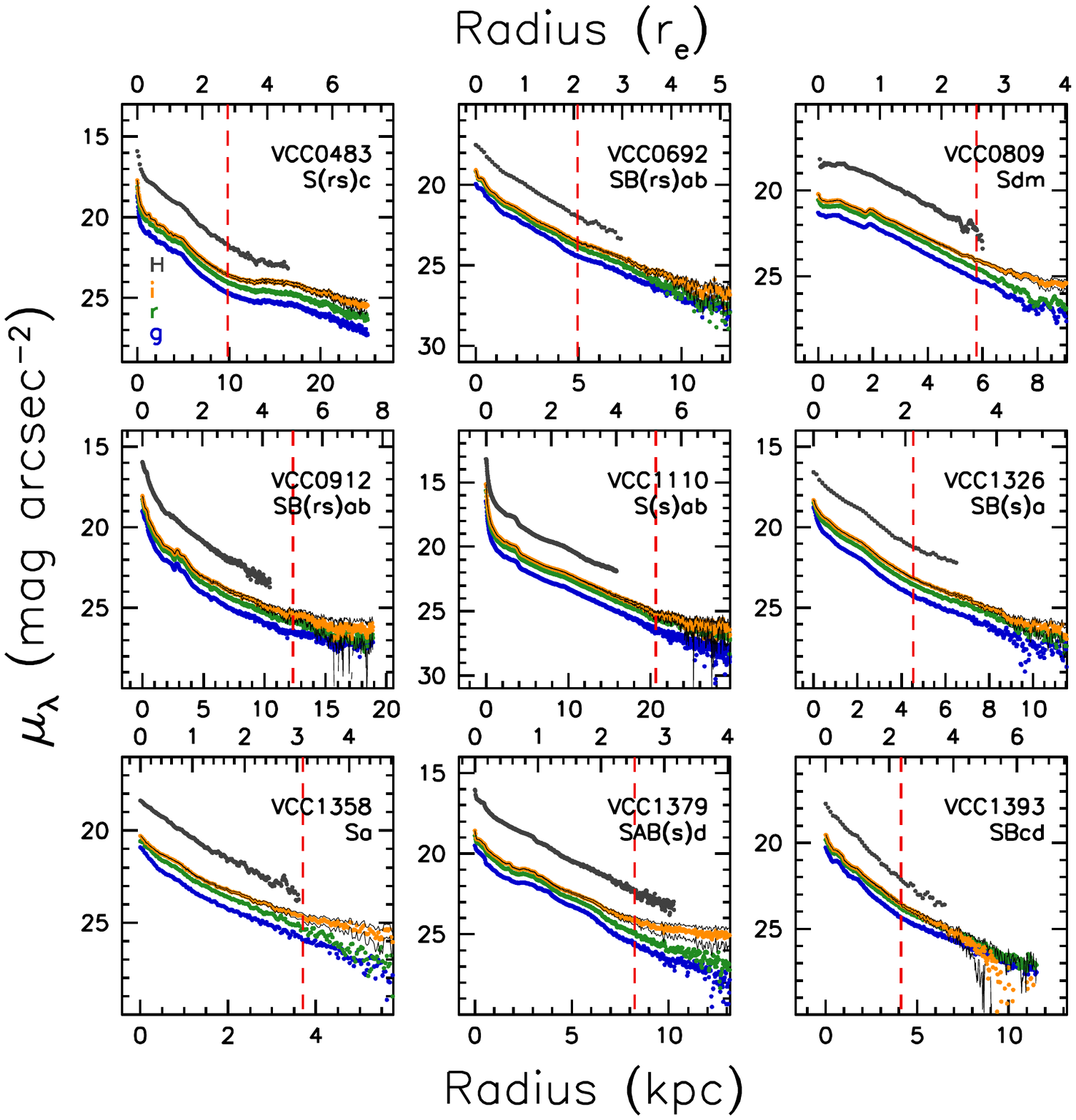}
  \caption{As in \Fig{SBPrfs-TypeI} but for Virgo Type III galaxies.  The 
location of the break for each galaxy is marked by the red vertical line.}
  \label{fig:SBPrfs-TypeIII}
 \end{center}
\end{figure*}

\clearpage
\begin{figure*}
 \ContinuedFloat
 \begin{center}
  \includegraphics[width=0.9\textwidth]{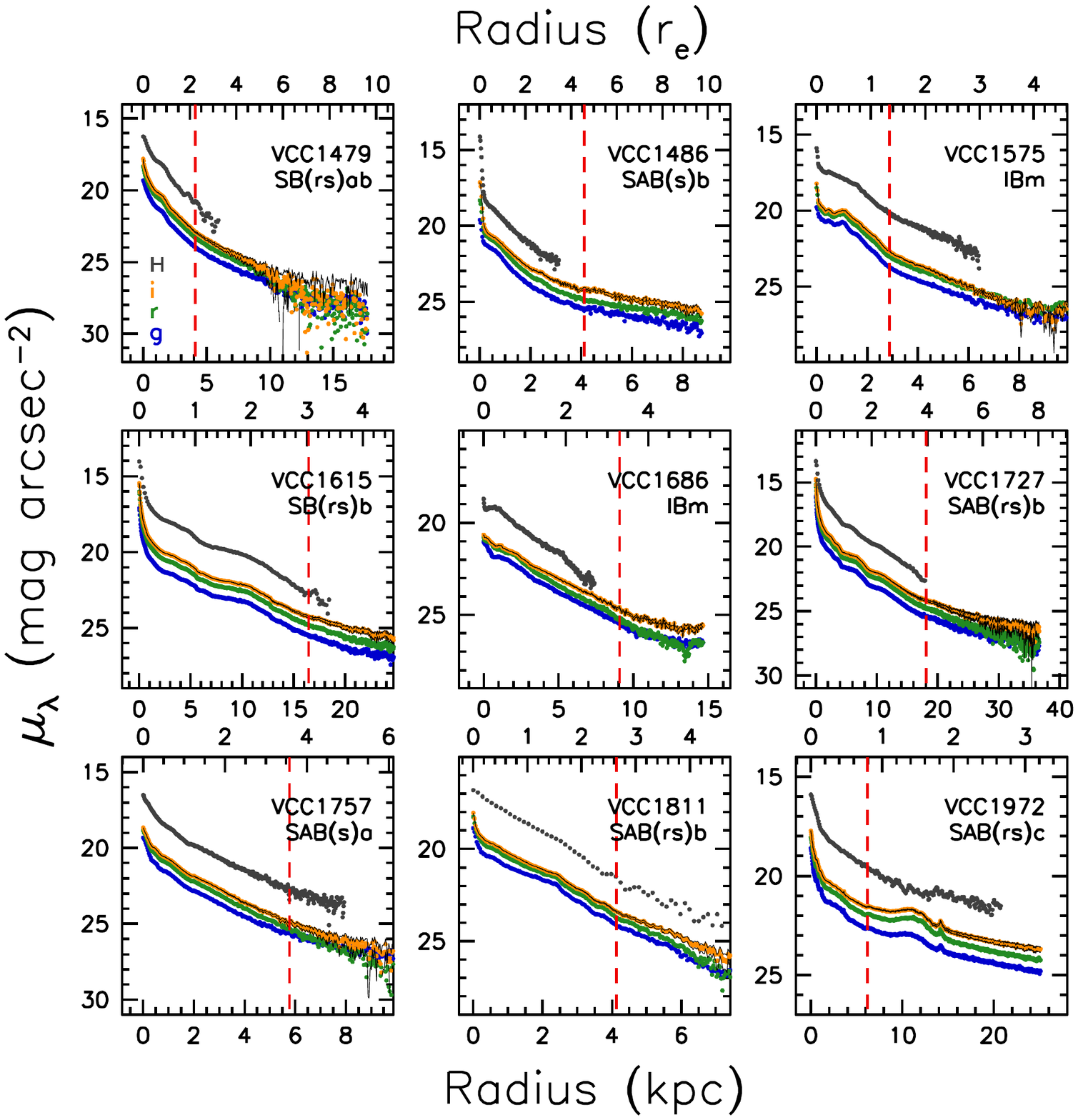}
  \caption{(\textit{continued})}
 \end{center}
\end{figure*}

\clearpage
\begin{figure*}
 \ContinuedFloat
 \begin{center}
  \includegraphics[width=0.9\textwidth]{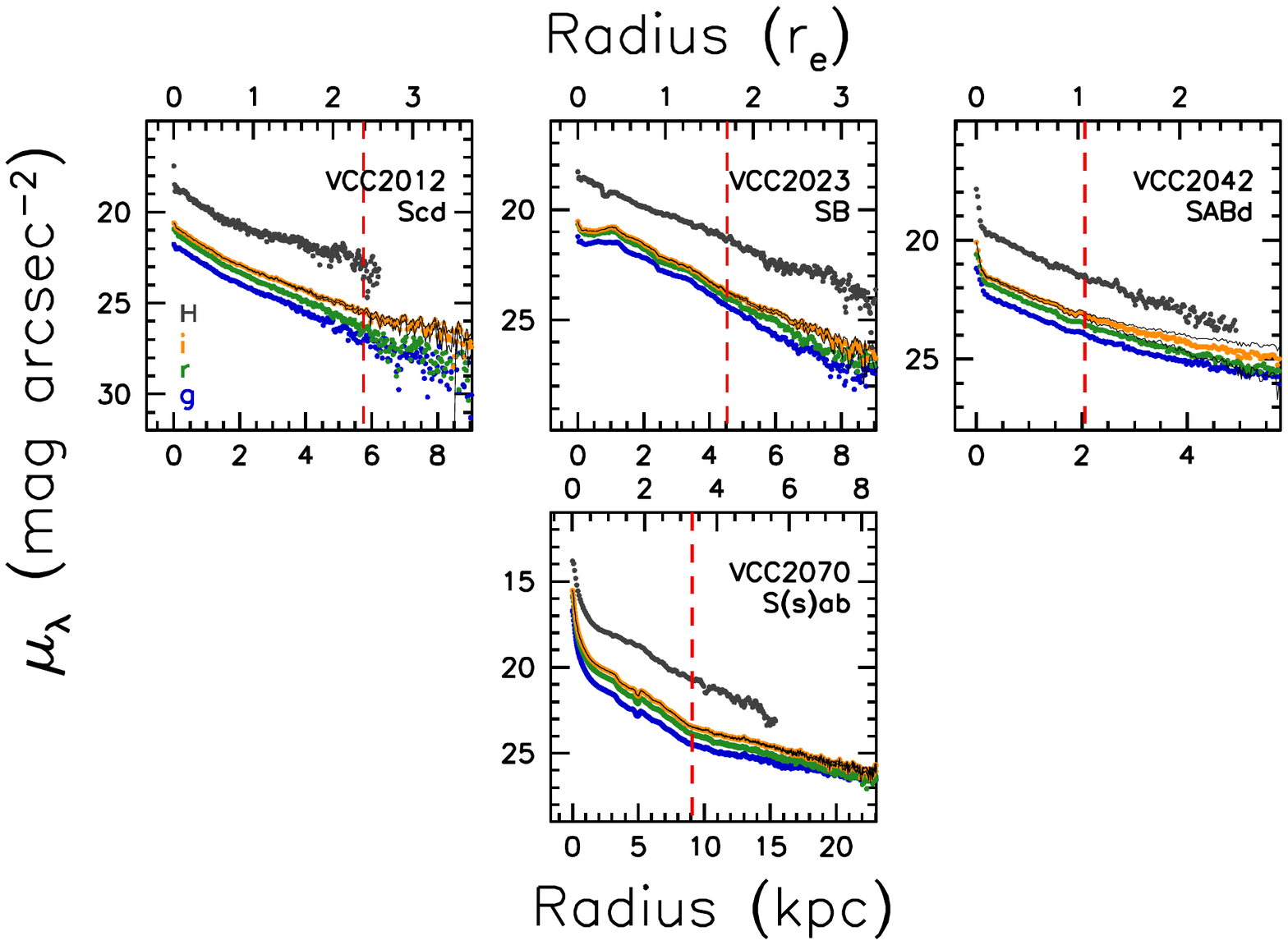}
  \caption{(\textit{continued})}
 \end{center}
\end{figure*}

\clearpage
\begin{figure*}
 \begin{center}
  \includegraphics[width=0.9\textwidth]{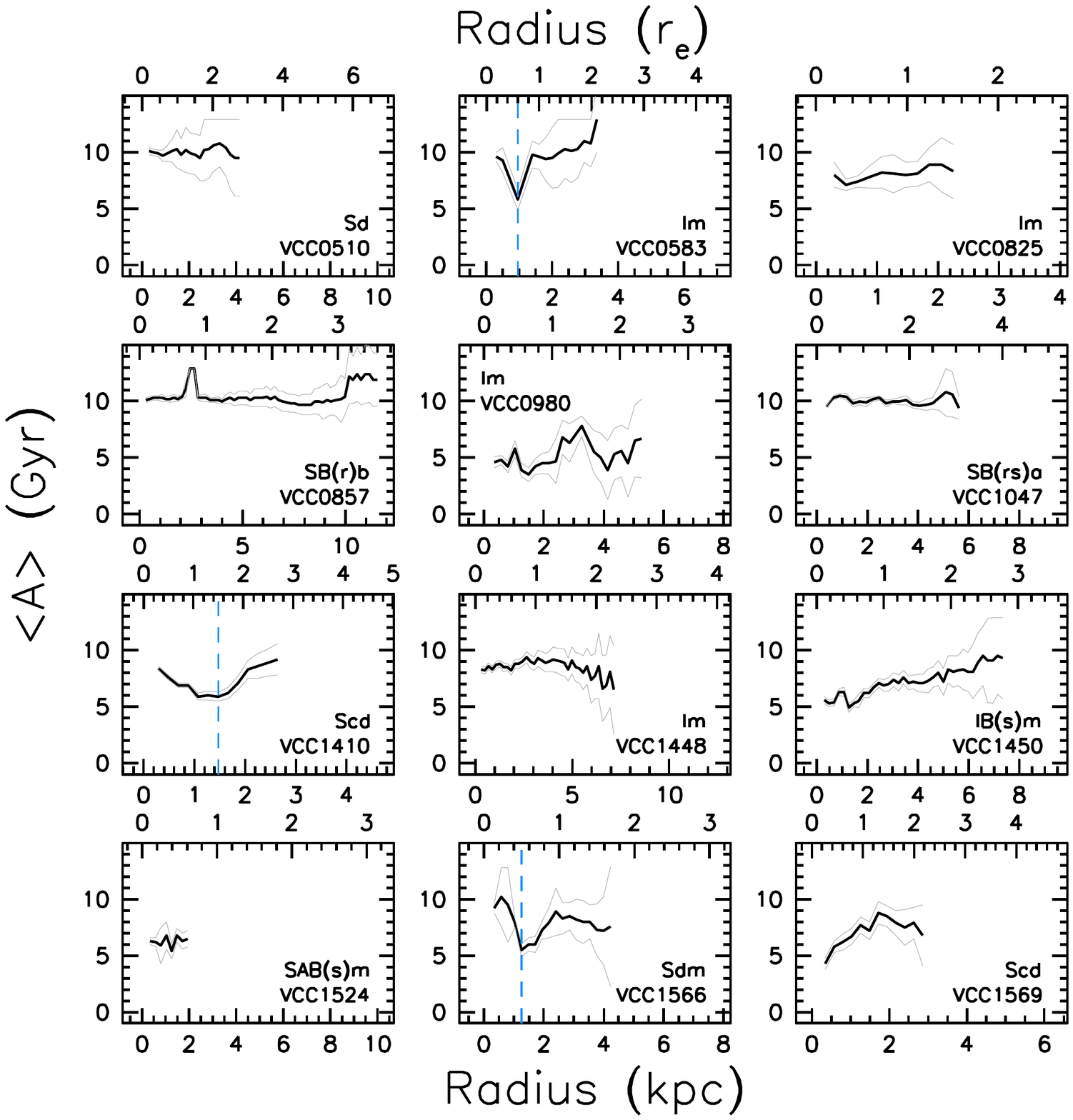}
  \caption{Mean stellar age profiles (black lines) and their 1$\sigma$ error 
envelopes (grey lines) for our sub-sample of Virgo Type I galaxies.  The VCC 
identification number and detailed morphology of each galaxy is provided in 
each panel, while the location of the age minimum (where applicable) is marked 
by the blue, dashed vertical line.  The ranges along the horizontal axes match 
those used in the corresponding panels of \Fig{SBPrfs-TypeI}.}
  \label{fig:APrfs-TypeI}
 \end{center}
\end{figure*}

\clearpage
\begin{figure*}
 \ContinuedFloat
 \begin{center}
  \includegraphics[width=0.9\textwidth]{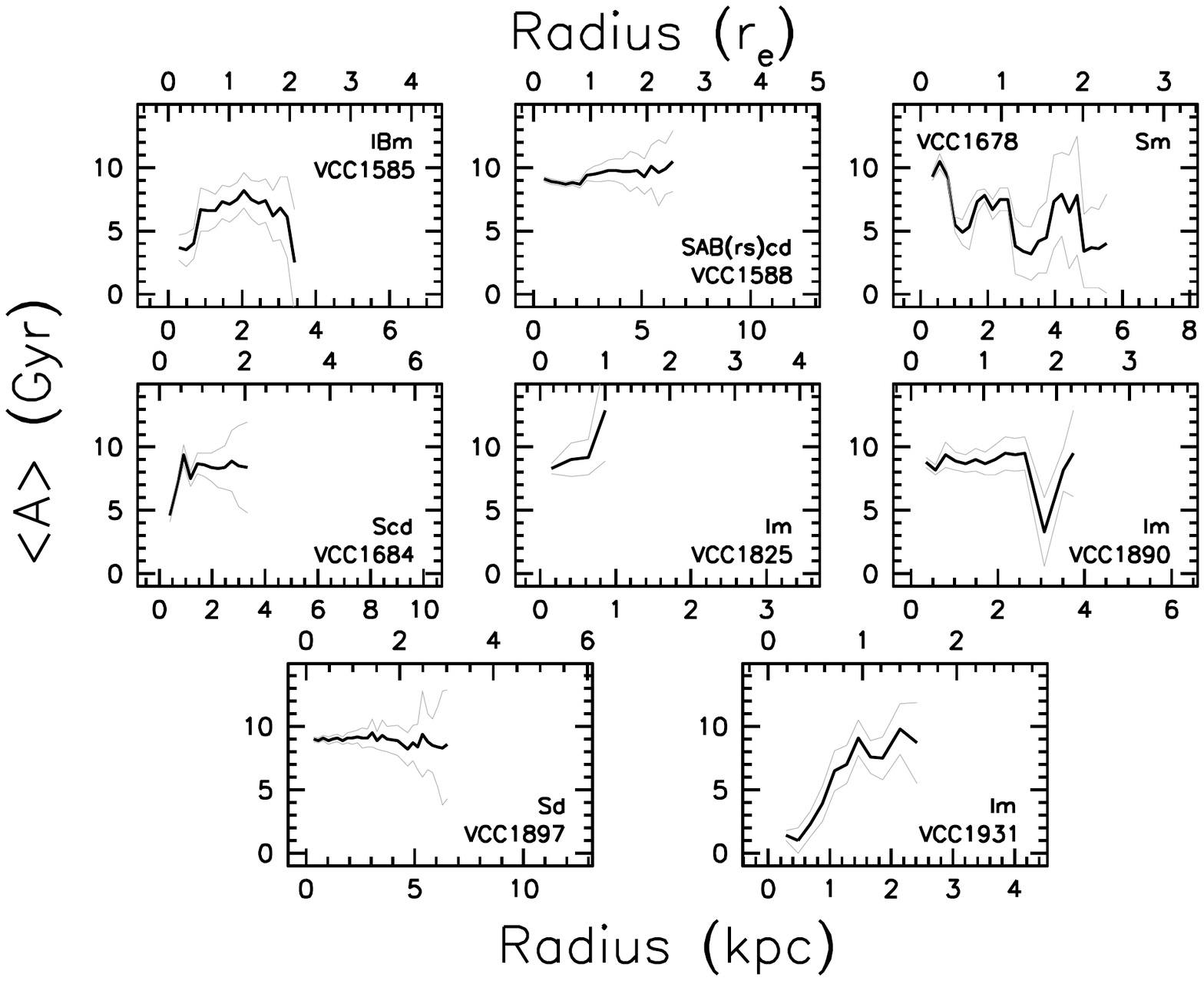}
  \caption{(\textit{continued})}
 \end{center}
\end{figure*}

\clearpage
\begin{figure*}
 \begin{center}
  \includegraphics[width=0.9\textwidth]{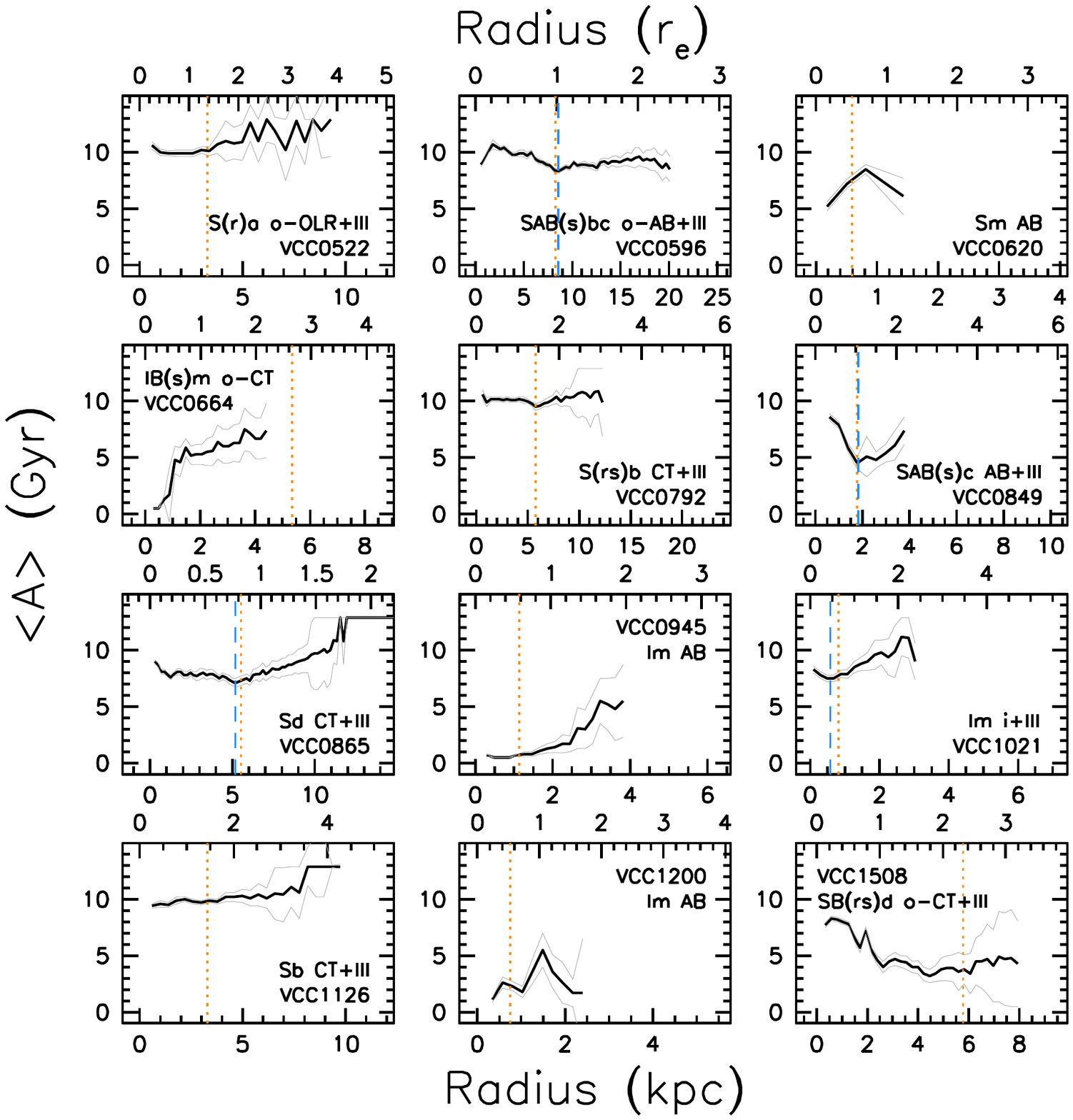}
  \caption{As in \Fig{APrfs-TypeI} but for Virgo Type II galaxies and now the 
ranges along the horizontal axes match those used in the corresponding panels 
of \Fig{SBPrfs-TypeII}.  The specific class of Type II break for each galaxy is 
also provided in each panel, while its location is marked by the orange 
vertical line.}
  \label{fig:APrfs-TypeII}
 \end{center}
\end{figure*}

\clearpage
\begin{figure*}
 \ContinuedFloat
 \begin{center}
  \includegraphics[width=0.9\textwidth]{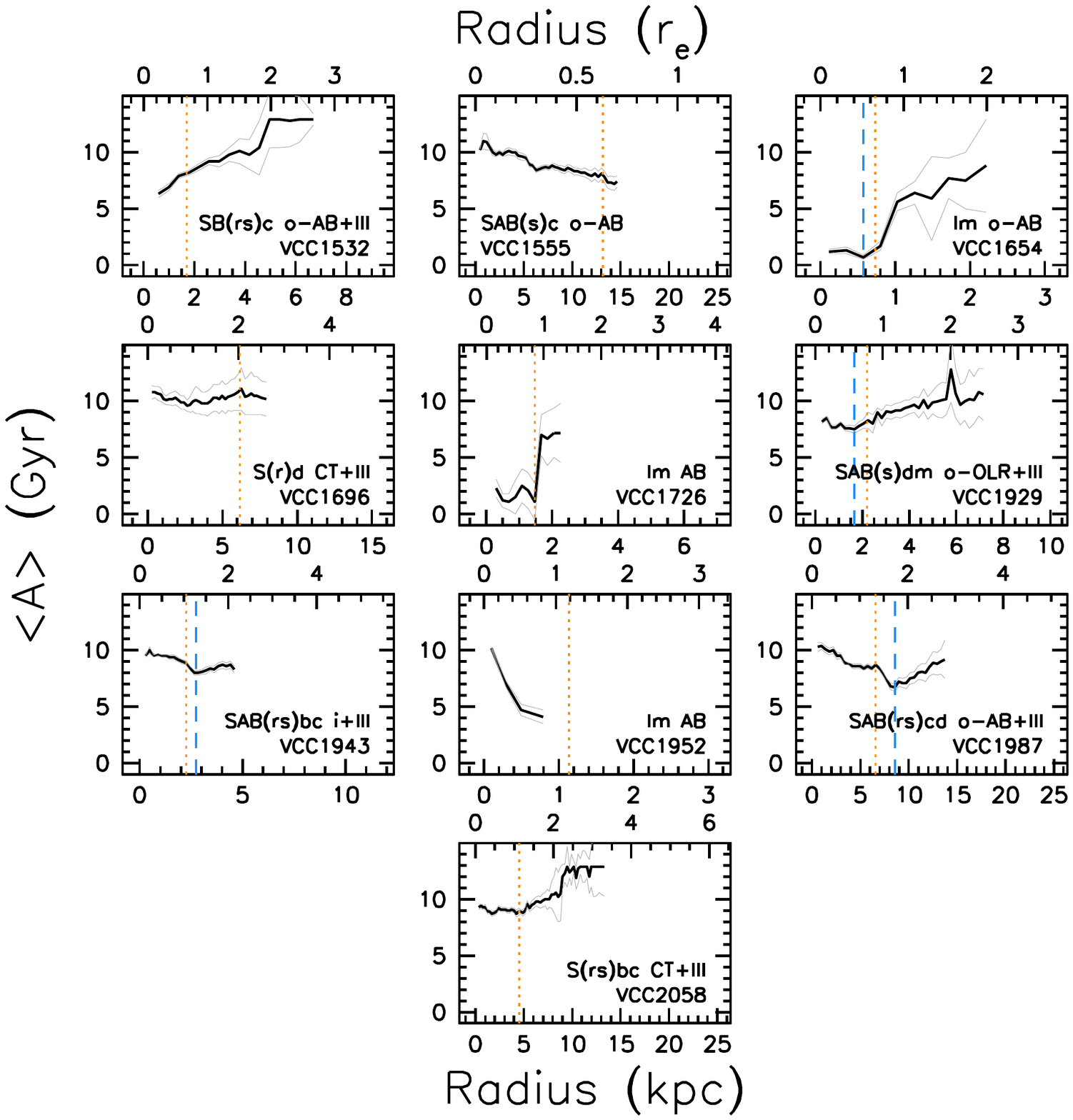}
  \caption{(\textit{continued})}
 \end{center}
\end{figure*}

\clearpage
\begin{figure*}
 \begin{center}
  \includegraphics[width=0.9\textwidth]{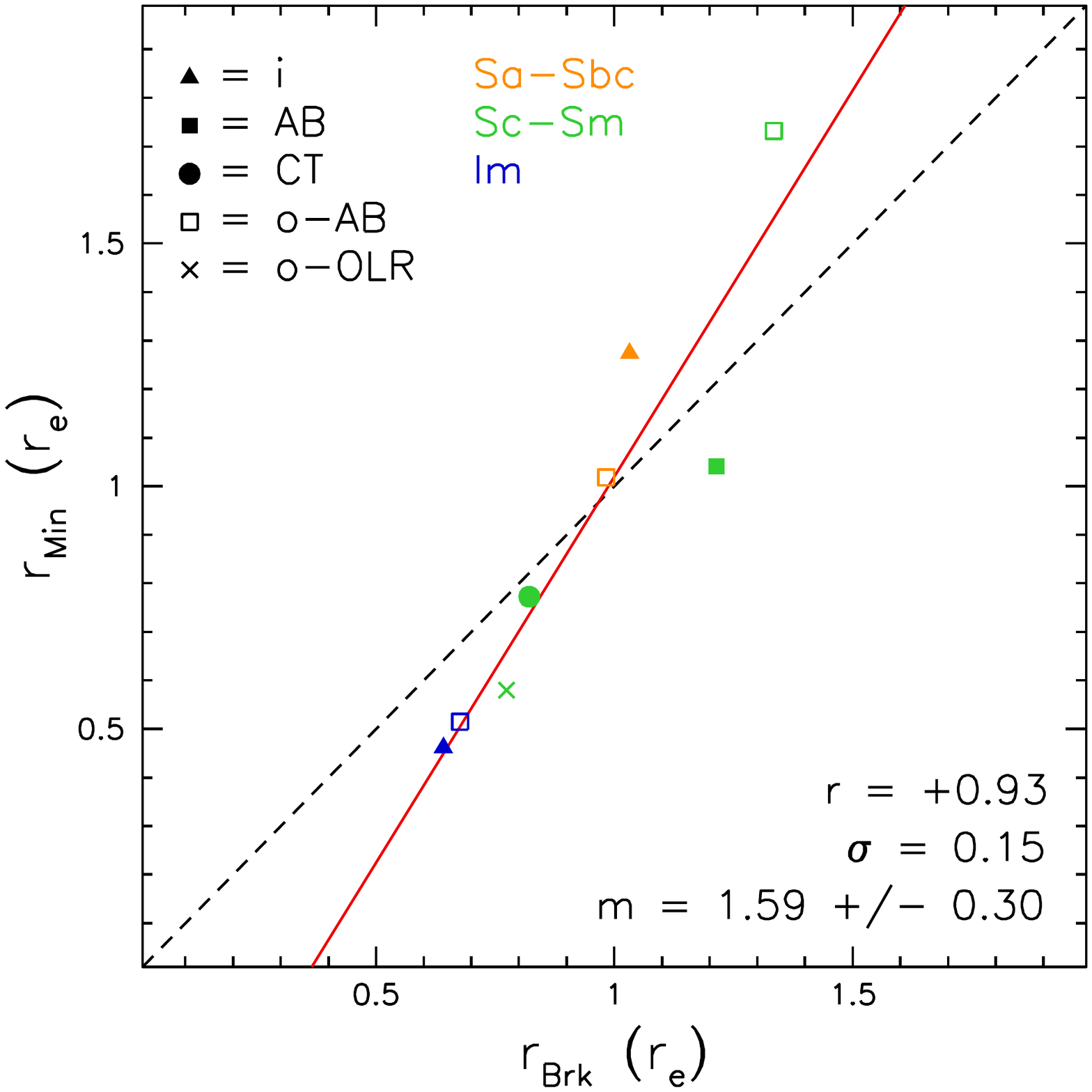}
  \caption{Location of age profile minima (r$_{\text{Min}}$) versus location of 
luminosity profile breaks (r$_{\text{Brk}}$) for those Virgo Type II galaxies 
exhibiting age profile inversions.  Both radii are scaled in terms of the 
galaxies' $H$-band effective radii (r$_{\text{e}}$).  The dashed line marks 
the 1:1 correlation.  The data points are coloured according to the galaxies' 
morphologies (Sa$-$Sbc = orange, Sc$-$Sm = green, Im = blue), while the point 
types reflect the Type II class (triangle = i, circle = CT, square = 
o-AB, cross = o-OLR).  The Pearson correlation coefficient, rms dispersion and 
bootstrap linear fit slope (red line) are also provided.}
  \label{fig:rMinvsrBrk}
 \end{center}
\end{figure*}

\clearpage
\begin{figure*}
 \begin{center}
  \includegraphics[width=0.9\textwidth]{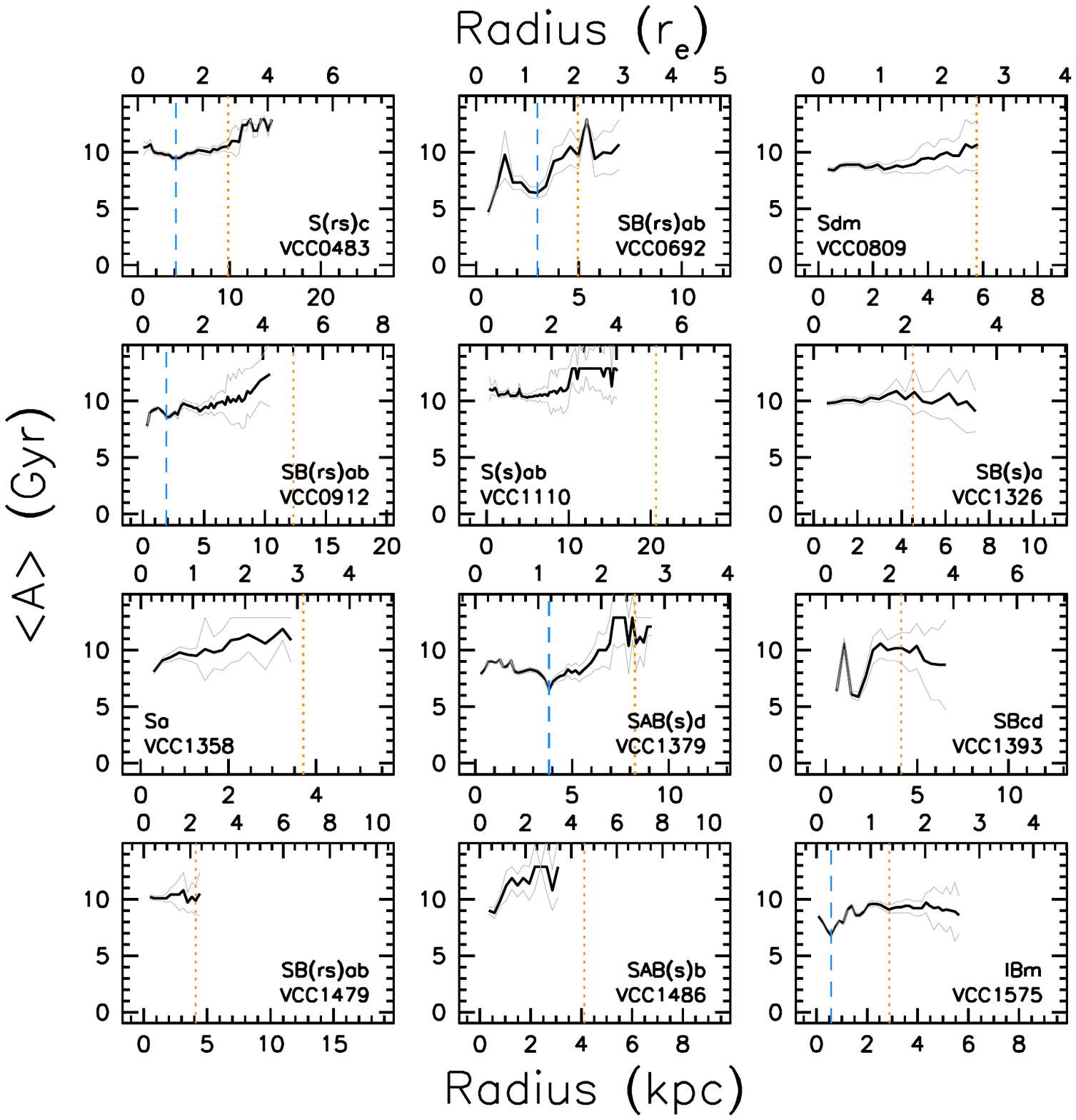}
  \caption{As in \Fig{APrfs-TypeI} but for Virgo Type III galaxies and now the 
ranges along the horizontal axes match those used in the corresponding panels 
of \Fig{SBPrfs-TypeIII}.  The location of the break for each galaxy is marked 
by the orange vertical line in each panel.}
  \label{fig:APrfs-TypeIII}
 \end{center}
\end{figure*}

\clearpage
\begin{figure*}
 \ContinuedFloat
 \begin{center}
  \includegraphics[width=0.9\textwidth]{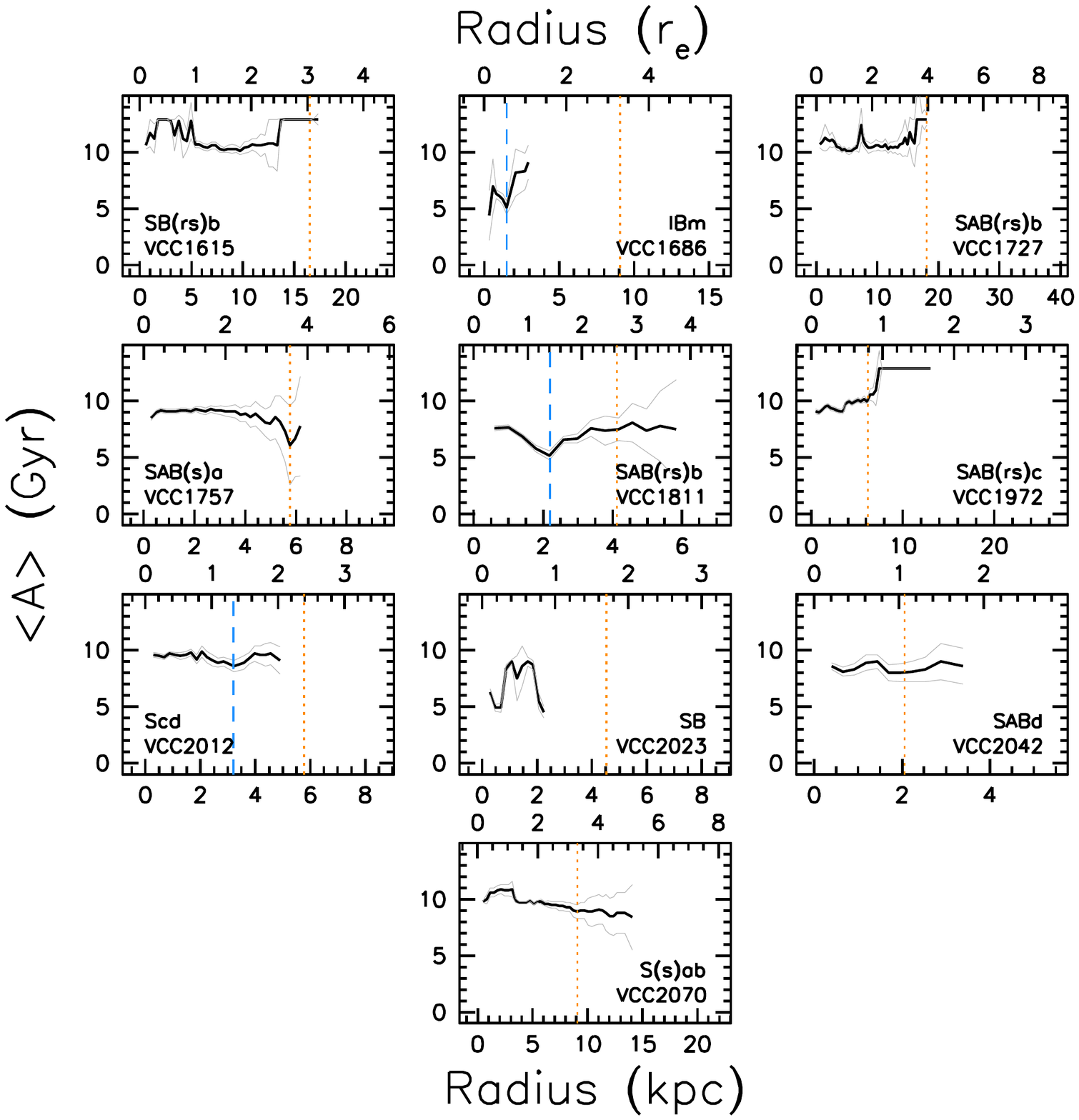}
  \caption{(\textit{continued})}
 \end{center}
\end{figure*}

\clearpage
\begin{figure*}
 \begin{center}
  \includegraphics[width=0.9\textwidth]{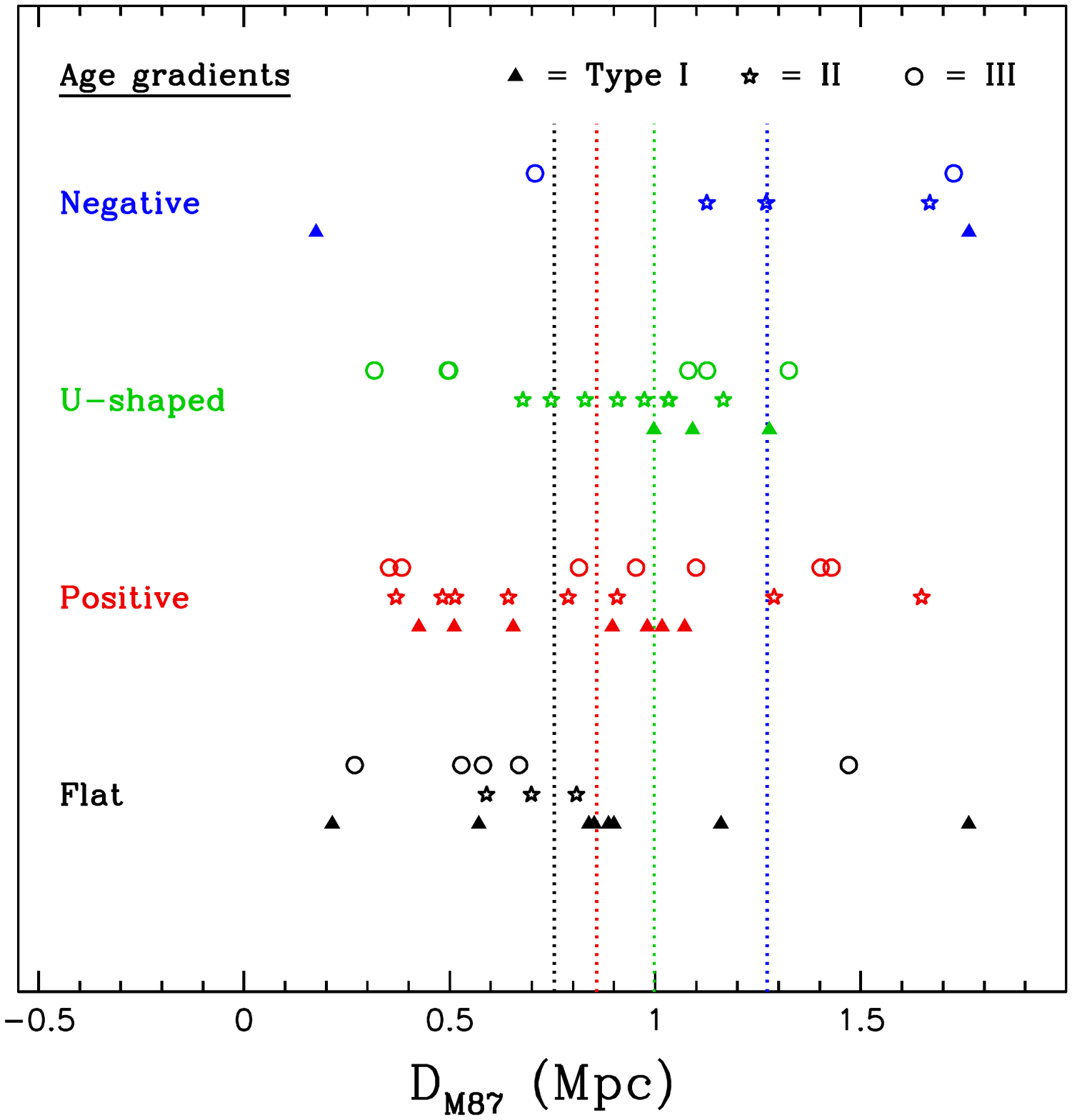}
  \caption{Differential distribution of (projected) cluster-centric radii for 
our Virgo disk galaxy sample.  Galaxies have been binned and coloured 
according to their age gradients (flat = black, positive = red, 
negative = blue, U-shaped = green).  The point types refer to their
luminosity profile shapes (Type I = triangle, II = star, III = circle).  The 
coloured vertical lines locate the median values of the four distributions.}
  \label{fig:DM87Distbn}
 \end{center}
\end{figure*}

\clearpage
\begin{figure*}
 \begin{center}
  \includegraphics[width=0.9\textwidth]{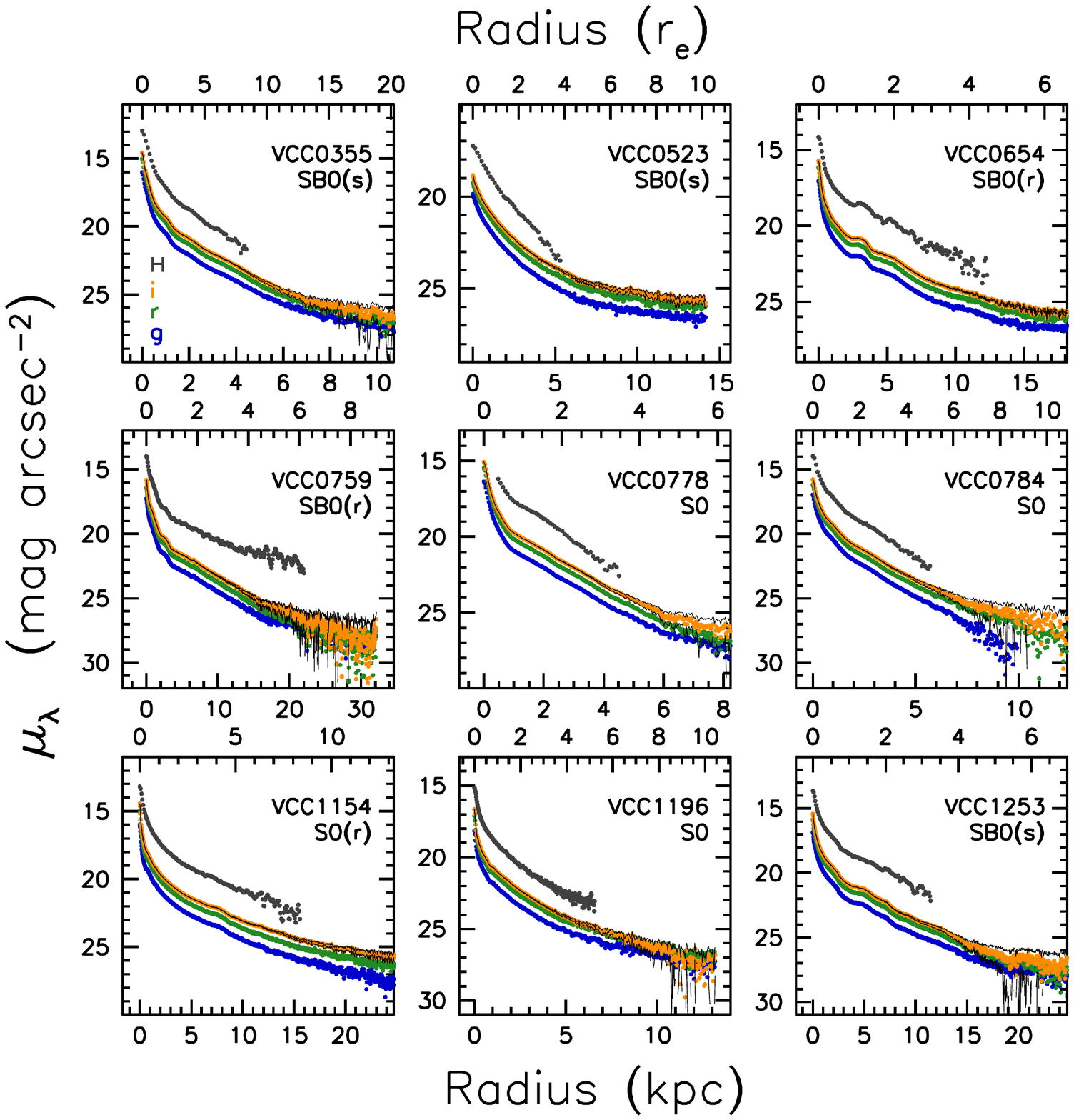}
  \caption{As in \Fig{SBPrfs-TypeI} but for Virgo S0s common to both our entire 
sample and that of \cite{Er12}.}
  \label{fig:SBPrfs-Er12}
 \end{center}
\end{figure*}

\clearpage
\begin{figure*}
 \ContinuedFloat
 \begin{center}
  \includegraphics[width=0.9\textwidth]{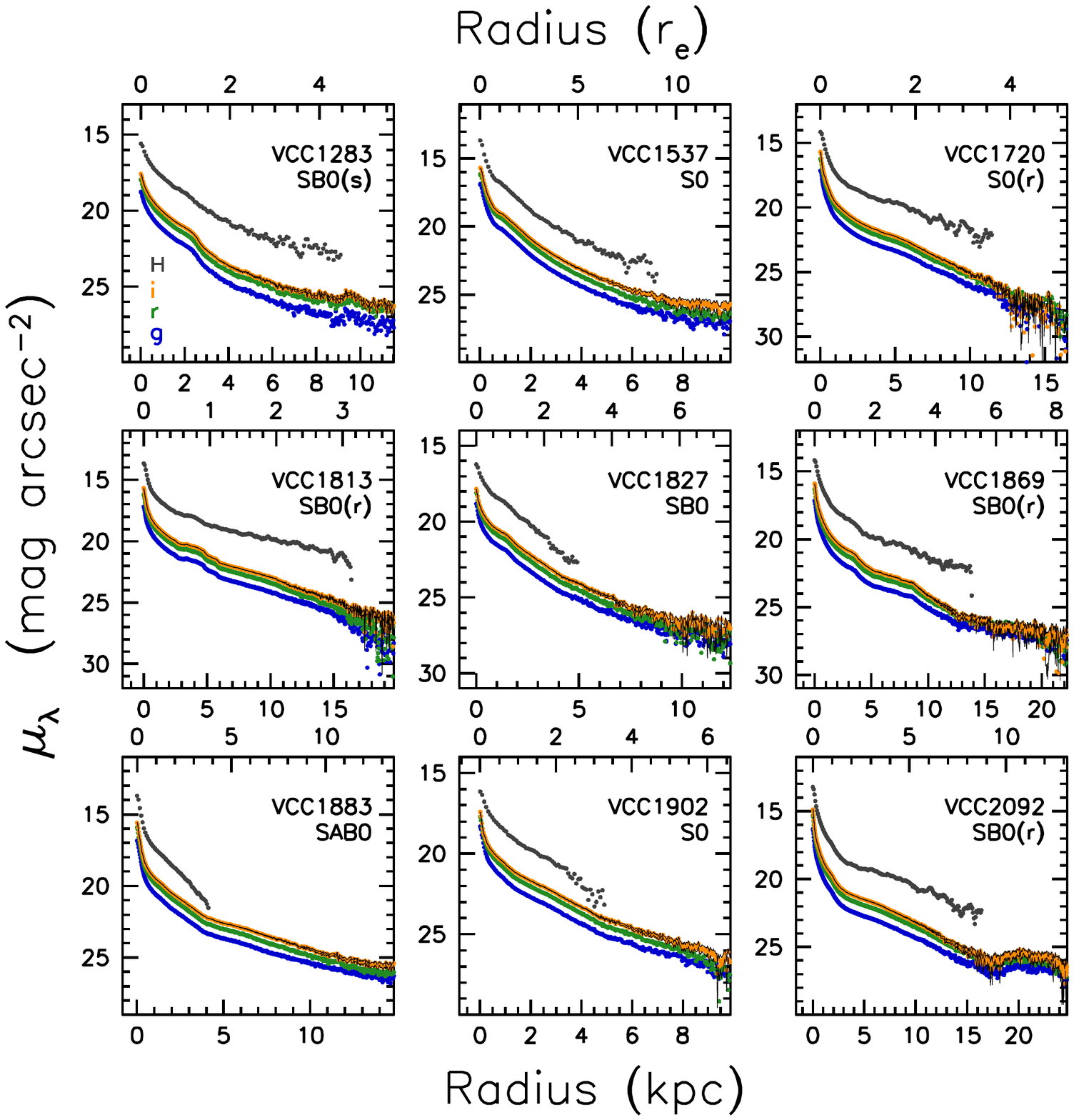}
  \caption{(\textit{continued})}
 \end{center}
\end{figure*}

\end{document}